\documentclass[sigconf]{acmart}
\renewcommand\footnotetextcopyrightpermission[1]{}

\usepackage{graphicx} 
\usepackage[ruled,vlined]{algorithm2e}
\usepackage{amsmath}
\usepackage{tikz}
\newcommand{\EbnoSmall}[0]{{{\mathcal{E}_b}/{N_0}}}

\newcommand{\Name}{StopSec}
\title[Sensing and Stopping Interfering Secondary Users]{\texorpdfstring{Sensing and Stopping Interfering Secondary Users:\\Validation of an Efficient Spectrum Sharing System}{Sensing and Stopping Interfering Secondary Users: Validation of an Efficient Spectrum Sharing System}}

\begin{document}

\author{Meles G. Weldegebriel}
\affiliation{
  \institution{Electrical \& Systems Engineering, Washington University in St. Louis}
  \city{St. Louis}
  \state{Missouri}
  \country{USA}
}
\author{Zihan Li}
\affiliation{
  \institution{Computer Science \& Engineering, Washington University in St. Louis}
  \city{St. Louis}
  \state{Missouri}
  \country{USA}
}
\author{Dustin Maas}
\affiliation{
  \institution{Kahlert School of Computing, University of Utah}
  \city{Salt Lake City}
  \state{Utah}
  \country{USA}
}
\author{Greg Hellbourg}
\affiliation{
  \institution{Cahill Center for Astronomy and Astrophysics, California Institute of Technology}
  \city{Pasadena}
  \state{California}
  \country{USA}
}
\author{Ning Zhang}
\affiliation{
  \institution{Computer Science \& Engineering, Washington University in St. Louis}
  \city{St. Louis}
  \state{Missouri}
  \country{USA}
}
\author{Neal Patwari}
\affiliation{
  \institution{Kahlert School of Computing, University of Utah}
  \city{Salt Lake City}
  \state{Utah}
  \country{USA}
}

\begin{abstract}
We present the design and validation of \textit{Stoppable Secondary Use (\Name{})}, a privacy-preserving protocol with the capability to identify a secondary user (SU) causing interference to a primary user (PU) and to act quickly to stop the interference. 
All users are served by a database that provides a feedback mechanism from a PU to an interfering SU. We introduce a new lightweight and robust method to watermark an SU's OFDM packet. Through extensive over-the-air real-time experiments, we evaluate  \Name{}  in terms of interference detection, identification, and  stopping latency, as well as impact on SUs. We show that the watermarking method avoids negative impact to the secondary data link, and is robust to real-world time-varying channels. Interfering SUs can be stopped in $<150$ ms, and when multiple users are simultaneously interfering, they can all be stopped.  Even when the interference is 10 dB lower than the noise power, \Name{} successfully stops interfering SUs within a few seconds of their appearance in the channel. \Name{} can be an effective spectrum sharing protocol for cases when interference to a PU must be quickly and automatically stopped. 
\end{abstract}

\maketitle

\thispagestyle{plain}  
\pagestyle{plain}

\section{Introduction}

Dynamic spectrum sharing is essential to maximizing the benefits of limited spectrum resources across diverse users~\cite{NSF2024spectrum}. However, the design of today's sharing systems can be made inefficient by designing for the \textit{worst-case} interference, to ensure entering secondary users (SUs) pose an extremely low risk to primary users (PUs). For instance, the U.S. Federal Communications Commission (FCC) opened the 6\,GHz band for secondary Wi-Fi use through an Automated Frequency Coordinator (AFC) framework and propagation model. This framework was challenged by incumbent microwave link operators like AT\&T~\cite{FCC2022AFC}, who argued that despite the conservative propagation model~\cite{Intel2024AFC}, interference ``inevitably'' occurs, and the lack of a mitigation mechanism made their microwave backhaul systems vulnerable. Similar disputes arose between the FCC and the Federal Aviation Administration (FAA) over C-band allocation, where worst-case modeling predicted interference to radar altimeters --- even though such interference was not observed in practice~\cite{RTCA2020FAA}. 

The challenge is that any channel model is imperfect, and sometimes will be far off. If we need to ensure that interference only exceeds a threshold 5\% of the time, we need $16\,\mathrm{dB}$ extra margin compared to the mean received interference power, assuming the channel model of~\cite{itur1997guidelines}. But if we can tolerate interference only 0.01\% of the time, we need $37\,\mathrm{dB}$ extra margin. The extra $21\,\mathrm{dB}$ margin results in the area of exclusion being multiplied by a factor of 25, or equivalently, 96\% of the possible sharing area being unused.  These examples illustrate how designing for worst-case scenarios can lead to overly conservative, inefficient spectral reuse.

This paper instead tests an alternate, reactive approach. Rather than using a very conservative model, we build a sharing system with stoppable SUs --- the ability to quickly and automatically stop a device from operating on the band when its signals actually are measured to interfere with a protected user. Manual enforcement mechanisms that do this are too slow — sometimes taking years to disable rogue transmitters~\cite{welch2014florida}. One existing reactive approach is to turn off subsets of SUs until the interference is no longer observed~\cite{sarbhai2024reactive}, but this can be slow, and some SUs are turned off when they are not interfering. Another idea is for SUs to watermark their RF transmissions to allow primary receivers to detect and identify the specific interfering user~\cite{Meles2024pseudonymetry, palacios2025hidden}. However, existing RF watermarking methods (1) can degrade the secondary wireless link by $3\,\mathrm{dB}$~\cite{Meles2024pseudonymetry}, (2) can add several ms of timing jitter~\cite{palacios2025hidden}, which is not acceptable for applications like 5G low latency, and (3) can fail in time-varying channels as exist in mobile wireless systems. To date, no published work has implemented RF watermarking over-the-air to stop interference in real time.

We propose, implement, and test \Name{} (\textit{Stoppable Secondary Use}), a protocol that enables real-time detection and stopping of the SU whose transmission is interfering in a spectrum sharing system. SUs in \Name{} pick pseudonyms and watermark their transmissions. If they interfere with a PU, even down to $10\,\mathrm{dB}$ below the noise floor, the pseudonym receiver decodes the pseudonyms, and logs them in a shared, time-aware database. SUs periodically query the database and switch band if their pseudonym has been reported. Our framework supports low-overhead coordination, minimizes false positives, and preserves SU privacy through unlinkable pseudonym generation.

Wireless channels are time-varying, and low-rate watermarking methods such as~\cite{Meles2022pseudonymetry} are sensitive to changes in the channel, the noise and interference. As we intend to detect interference that is on the order of $10$ dB below the noise, even small channel changes are problematic. We solve this by using a time-varying code in the pulse shape of the watermarking modulation that effectively cancels out the channel changes, on average, and focuses on the watermark. Further, the modulation is designed to make reception simple --- the watermark receiver uses energy detection, and no synchronization (symbol or frequency) is required. This is a critical requirement for reliable operation when the interfering signal is so low in signal-to-noise ratio (SNR).

Further, we address the problem that a watermark can degrade an SU communication link. In \Name{}, which is designed for multicarrier modulated signals, we chose to encode pseudonym bits in just one of the subcarriers, in contrast to modifying the entire SU communication signal as in \cite{Meles2024pseudonymetry,palacios2025hidden}. The impact is comparable to a pilot channel, which is 1.5\% or less of the bandwidth of the SU signal in our experiments. 

We implement \Name{} and perform extensive lab and over-the-air experiments to quantify and validate its effectiveness.  We quantitatively validate our new watermarking scheme in comparison with related work. We show that by using a small percentage of the bandwidth for pseudonym communication, \Name{} operates to shut off an interfering SU's transmissions in less than 650 ms when the interfering signal is at $-10\,\mathrm{dB}$ SNR, and less than 270 ms when it is $\ge -6\,\mathrm{dB}$. We also show that \Name{} shuts off interferers whose transmissions overlap in time, the first such demonstration of this capability.  Our system and experimental code~\cite{stopsec2025} is public and can be rerun on the POWDER wireless testbed.


\section{System Overview} \label{sec:overview}

\begin{figure}[tbhp]
  \centering
  \includegraphics[width=0.9\linewidth]{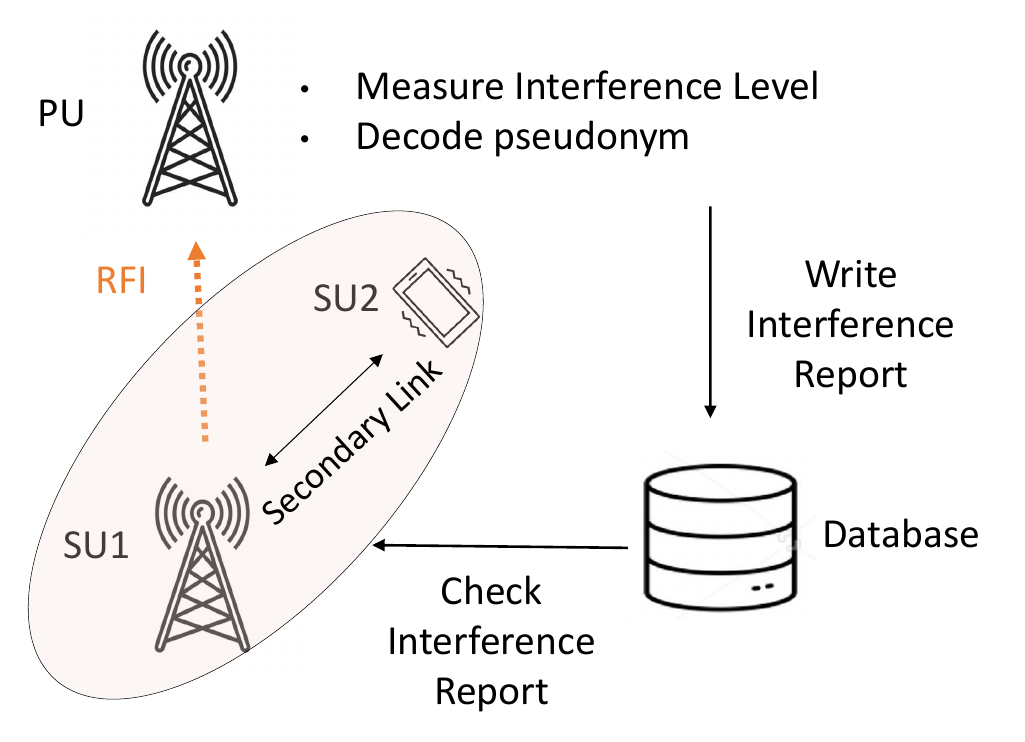}
  \caption{\Name{} system model.}
  \label{fig:sys_over}
\end{figure}

As shown in Figure~\ref{fig:sys_over}, \Name{} is a spectrum sharing system with (at least) two classes of users, primary and secondary, both connected to an interference report database. This section provides an overview, presenting the basic assumptions and actors, so that we can discuss the system tradeoffs and design choices in the following section.

\Name{} can operate as a standalone new protocol (as we evaluate in this paper), or an extension to existing protocols. In either case, SUs and PUs must take on the roles described in this section. 
\Name{} sets a standard watermarking method, which is known to both SUs and PUs. The watermark is designed to minimally impact the performance of the secondary communication system. The watermark is low in data rate and is able to be reliably decoded by primary receivers even at low received power.

The system operates with a closed-loop control mechanism shown in Figure~\ref{fig:sys_over}, requiring both PUs and SUs to have a secondary communication channel to access a remote database. Since randomly SU-generated pseudonyms do not translate to a static ID or address for an SU, a PU cannot use it to directly communicate with the SU. Instead, the control mechanism is implemented via a shared database.

In summary, an SU is permitted to transmit in a shared band only if the database --- which contains any interference reports from PUs --- confirms that the SU is not currently causing interference.

\subsection{Secondary User}
\label{subsectionTX}

We assume an Orthogonal Frequency Division Multiplexing (OFDM)-based SU communications system that is sharing spectrum with, and operating with lower priority than, the PU system. To enable \Name{}, the SU does two things:
\begin{enumerate}
    \item SUs are cooperative in the sense that they will vacate a shared band as needed to protect PUs. An SU periodically checks if it is currently interfering with operating PUs by querying the database. If it finds its pseudonym with matching metadata, it switches band. If not, it continues as planned. 
    \item The SU watermarks its data packet transmissions (called the host signal) with a self-generated random code called a pseudonym. The SU records locally its used pseudonym and the time and channel on which it was used. The watermark modulation and demodulation is described in Section \ref{subsec:watermarking}.
\end{enumerate}

\subsection{Primary User}
\label{subsectionRX}
\Name{} imposes an additional role on the PU: when interference is observed, it demodulates the received pseudonym from the interference signal, and writes an \textit{interference report} containing the pseudonym and metadata (timestamp, location, and frequency band) to the database described in Section~\ref{subsectionDB}.  The PU writes the interference report only if the pseudonym is correctly demodulated, which in effect ensures an SU is reported only when its interfering power is above a power threshold, e.g., -10 dB compared to the noise power in our experiments. 

\subsection{Database System}\label{subsectionDB}

The database serves as a repository for storing recent interference reports sent by PUs. It provides a feedback loop from a PU to the SU that is responsible for causing the interference. In this paper, we assume that the SU exclusively performs read operations, retrieving interference data necessary for changing its transmission parameters (e.g., channel), while the PU performs write operations, reporting instances of interference. The database is designed for security and low latency, which is described further in the following section.


\section{System Design}
\label{subsec:system design}

Given the key parts of \Name{} as described in Section \ref{sec:overview}, the key question is how to design each component to achieve a privacy-preserving, efficient, reliable and secure system to automatically stop interfering SUs. We address this here in this section by addressing key cross-cutting design issues: scalability, security and privacy, watermarking robustness, pseudonym packet design, detection, and false positive reduction.

We begin by detailing the core design elements of \Name{} --- its watermarking framework, pseudonym generation and management, and pseudonym detection mechanism at the PU. We then describe the interference database architecture and explain how it enables scalable, low-latency coordination across distributed wireless deployments. Finally, we describe techniques used to preserve user anonymity and system integrity.

\subsection{Watermarking Design}
\label{subsec:watermarking}
\noindent 
In this section, we introduce two watermarking strategies used in \Name{} focusing on improved pseudonym detection and impact of watermarking on data demodulation:
\begin{enumerate}
    \item \textbf{Coded modulation}. The wireless channel is time-varying, 
    and channel induced changes could overwhelm the watermark. Our observation is that if we code the watermark pulse shape, i.e., make it change over time, we can decode it by observing how much measured changes match the code.
    \item \textbf{Single subcarrier}. Narrowing the bandwidth is a key to increasing reliability at low received powers, as has been observed in low-power wide area networking~\cite{yang2017narrowband,vejlgaard2017coverage}. Existing RF watermarking methods encode the watermark across the entire host signal bandwidth~\cite{Meles2024pseudonymetry,palacios2025hidden}.  By matching the watermark data rate to its RF bandwidth, we efficiently encode bits with minimal impact to the SU communication system.
\end{enumerate}

\subsubsection{Coded Modulation}
\label{subsec:pattern}\

\noindent In spectrum sharing systems, PUs often must be protected from interference even lower in power than the noise floor. For example, in the 6\,GHz band, the AFC must ensure that interference power is at $-6\,\mathrm{dB}$ compared to the noise~\cite{FCC2022AFC,dogantusha2025evaluation}. At such levels, the SU's packet data is too weak to be correctly demodulated or excised by interference removal methods. The goal of \Name{} is to provide the primary receivers the means to identify and stop interference even at such low SNRs. 

Coded modulation can be contrasted with pulse amplitude modulation (PAM)~\cite{Meles2022pseudonymetry}. 
In PAM watermarking, the data signal is transmitted at higher amplitude when watermarking with a pseudonym bit 1 and at lower amplitude when watermarking with a pseudonym bit 0. In coded modulation (CM), a pseudonym bit is transmitted as a sequence of higher and lower amplitudes. In it, the pseudonym symbol period, $N$, is divided into $L$ chips, each of duration $N/L$. Each chip has a different (but known) amplitude. The pseudonym symbol becomes a sequence of known unique amplitudes, i.e., the \textit{code}. CM uses a higher rate unique and known code during each symbol period. However, in CM, the amplitudes are all non-negative so that the receiver can use the instantaneous received power to decode the watermark. Energy detection avoids phase synchronization, which fails at low SNR~\cite{rice2008digital}. 
Codes can be generated using maximum-length pseduo-noise sequences (m-sequences), can be alternate between two amplitudes, or other pattern.

\begin{figure}[tbhp]
  \centering
    \includegraphics[width=0.9\linewidth]{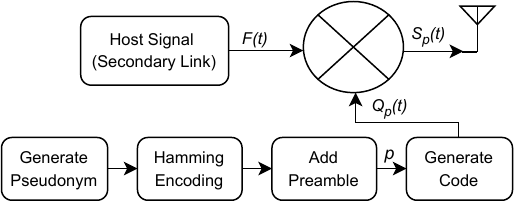}
      \caption{Framework for CM watermarking.}
     \label{fig:watermark-generation}
\end{figure}

Figure~\ref{fig:watermark-generation} presents the watermarking framework. To create the CM signal, we create one pseudonym bit symbol $Q_p(t)$ as,
\begin{equation} 
\label{E:watermark}
    Q_p(t) = \sum_{l=0}^{L-1}  \left[1-A_p[l]  \alpha \right]\phi(t -lT_c),
\end{equation}
where $p \in \{0,1\}$ is the pseudonym bit to send, $l$ is the chip number, $A_p[l]$ is amplitude of chip $l$ for bit $p$, $\alpha$ is the modulation index, $T_c$ is the chip duration, and $\phi(t)$ is the chip pulse shape. To simplify the notation used throughout the rest of this paper, we use ``P-bit'' to refer to the pseudonym symbol. Since we send one bit per P-bit in all of our experiments, there is no difference between pseudonym symbol and bit. 

Figure~\ref{fig:watermark-generation} illustrates the generation of watermarked signal in CM. A unique code $Q_p(t)$ is generated for each p-bit (0 or 1) and is used to modulate the host signal $f(t)$, resulting in the watermarked signal $S_p(t)$. The watermarking operation is defined as $S_p(t)= Q_p(t)f(t)$. In order to allow us to evaluate CM separately from other \Name{} strategies, we consider two cases for the host signal $f(t)$. When we refer to CM alone, $f(t)$ corresponds to the entire data packet from the SU communication link. In the following section, we consider when $f(t)$ is just one of the subcarriers of an SU's OFDM signal.

\subsubsection{Single Subcarrier Pseudonym Communication}
\label{subsub:single-subcarrier}\

\noindent A key goal in \Name{} is to watermark without significant impact to the SU communication system. We provide an alternative to existing watermark methods: the PAM watermarking scheme in~\cite{Meles2022pseudonymetry} degrades data demodulation performance at secondary receivers by $1-3$ dB depending on the modulation index, and the pulse position modulation scheme in~\cite{palacios2025hidden} adds timing jitter into the communication link. Even though these modifications may be a reasonable tradeoff in particular application scenarios, it is critical to minimize the impact of watermarking on the SU's communication link. In multi-carrier communication systems like OFDM, we propose to reserve one of the subcarriers to send the watermark signal. This is a dedicated `pseudonym subcarrier', not modulated by regular SU communication data. Our assumption is that in a cooperative spectrum sharing, dedicating one out of all subcarriers for pseudonym communication is a small investment. For example, this corresponds to 1 out of 256 subcarriers in IEEE 802.11ax (Wi-Fi 6). OFDM subcarriers are used for link reliability purposes, e.g., several subcarriers are used as pilot channels, and several more are used as guard band subcarriers. As our subcarrier is essentially a pilot that turns on and off, it is similar to a pilot subchannel. And further, because it changes at a low rate compared to the symbol rate, it has lower sidelobes in the frequency domain than a data subcarrier, and thus could replace a guard band subcarrier.

To enable reliable low-SNR detection on the pseudonym subcarrier, we apply CM watermarking, which we refer to together as \Name{} watermarking. This corresponds to Figure~\ref{fig:watermark-generation}, but with the pseudonym subcarrier as $f(t)$ --- only it is watermarked. This means that the other subcarriers remain intact, and there is no performance impact to data demodulation at SU receivers. Further, we use 100\% modulation on the pseudonym subcarrier, i.e., $\alpha=1$ in (\ref{E:watermark}). Either the subcarrier is totally off, or it is a carrier transmitted with twice the amplitude of the other subcarriers.
We also study the performance of \Name{} in data demodulation on the secondary link (Figure~\ref{fig:sys_over}). Experimental results for \Name{} are described in detail in the Section~\ref{Evaluation}.

\subsection{Pseudonym Packet: Management}

In the prior section, we describe how to watermark a bit onto an SU's packet. In this section, we describe how an entire (many bit) pseudonym is sent over multiple SU packets, and how the process is made privacy-preserving and reliable.

In \Name{}, SUs self-generate randomized pseudonyms to use to watermark their transmissions, without revealing their identity. A PU passively demodulates these pseudonyms if the SU's interference power is high enough for the watermark to be successfully demodulated, and logs them with associated timestamps and channels in a shared database. Before any transmission, an SU queries this database: if its pseudonym is found, it infers it is currently interfering, and does not transmit on the same band to avoid interfering with the PU. To make this mechanism reliable, scalable, and privacy-compliant, the system integrates a set of complementary design components.

\subsection{Pseudonym Packet Structure}
Figure 3 shows the frame structure for a pseudonym packet, which consists of two fields: a 7-bit preamble and a coded pseudonym. The preamble is a maximal-length sequence (m-sequence) \cite{GolombMLS}, selected for its strong autocorrelation properties. It marks the start of the frame and enables the receiver to synchronize and filter out corrupted or spurious transmissions.

The coded pseudonym follows the preamble and represents the SU’s temporary identity in a privacy-preserving manner. Each pseudonym comprises 26 random bits, encoded using a (31, 26) Hamming code~\cite{lin2004error}. This forward error-correcting code supports single-bit correction and double-bit error detection, adding robustness to pseudonym detection capability.

The use of a 26-bit pseudonym has a balance between scalability and reliability. Specifically, it limits the probability of false positives during suppression. If a pseudonym consists of $K$ random bits and only one SU is transmitting, the probability that a second SU coincidentally uses the same pseudonym --- at the same time, on the same channel is $2^{-K}$. In our case, this corresponds to a false alarm probability of $2^{-26} \approx 1.5 \times 10^{-8}$ --- a negligible risk even in dense deployments.

At the PU, pseudonym packets are accepted only if they pass two integrity checks:
\begin{enumerate}
    \item \textbf{Preamble Correlation}. The receiver verifies that the initial 7 bits match the expected m-sequence. Frames failing this check are discarded.

    \item \textbf{Error Control Coding}. Hamming decoding corrects and validates the encoded pseudonym bits. Only validated frames lead to writing to the interference database.
\end{enumerate}
These layered validation steps ensure high pseudonym detection accuracy, even under low-SNR conditions or when multiple SUs transmit concurrently.

\begin{figure}[tbp]
  \centering
      \tikzset{every picture/.style={line width=0.75pt}} 

\begin{tikzpicture}[x=0.75pt,y=0.75pt,yscale=-1,xscale=1]

\draw   (104.5,128) -- (201.5,128) -- (201.5,152) -- (104.5,152) -- cycle ;
\draw   (201.5,128) -- (423.5,128) -- (423.5,152) -- (201.5,152) -- cycle ;
\draw    (121.5,117) -- (104.5,117) ;
\draw [shift={(104.5,117)}, rotate = 360] [color={rgb, 255:red, 0; green, 0; blue, 0 }  ][line width=0.75]    (0,5.59) -- (0,-5.59)(10.93,-4.9) .. controls (6.95,-2.3) and (3.31,-0.67) .. (0,0) .. controls (3.31,0.67) and (6.95,2.3) .. (10.93,4.9)   ;
\draw    (182.5,116) -- (201.5,116) ;
\draw [shift={(201.5,116)}, rotate = 180] [color={rgb, 255:red, 0; green, 0; blue, 0 }  ][line width=0.75]    (0,5.59) -- (0,-5.59)(10.93,-4.9) .. controls (6.95,-2.3) and (3.31,-0.67) .. (0,0) .. controls (3.31,0.67) and (6.95,2.3) .. (10.93,4.9)   ;
\draw    (353.5,116) -- (423.5,116) ;
\draw [shift={(423.5,116)}, rotate = 180] [color={rgb, 255:red, 0; green, 0; blue, 0 }  ][line width=0.75]    (0,5.59) -- (0,-5.59)(10.93,-4.9) .. controls (6.95,-2.3) and (3.31,-0.67) .. (0,0) .. controls (3.31,0.67) and (6.95,2.3) .. (10.93,4.9)   ;
\draw    (276.5,116) -- (201.5,116) ;
\draw [shift={(201.5,116)}, rotate = 360] [color={rgb, 255:red, 0; green, 0; blue, 0 }  ][line width=0.75]    (0,5.59) -- (0,-5.59)(10.93,-4.9) .. controls (6.95,-2.3) and (3.31,-0.67) .. (0,0) .. controls (3.31,0.67) and (6.95,2.3) .. (10.93,4.9)   ;

\draw (123,135) node [anchor=north west][inner sep=0.75pt]   [align=left] {{\fontfamily{pcr}\selectfont Preamble}};
\draw (217,135) node [anchor=north west][inner sep=0.75pt]   [align=left] {{\fontfamily{pcr}\selectfont Hamming encoded pseudonym}};
\draw (130,109) node [anchor=north west][inner sep=0.75pt]   [align=left] {{\fontfamily{pcr}\selectfont 7 bits}};
\draw (287,109) node [anchor=north west][inner sep=0.75pt]   [align=left] {{\fontfamily{pcr}\selectfont 31 bits}};

\end{tikzpicture}
      \caption{Frame structure for pseudonym packets.}
     \label{fig:frame-structure}
\end{figure}
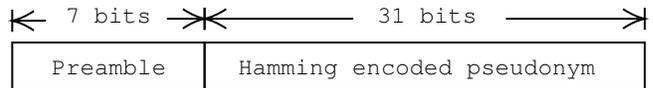

\subsubsection{Privacy and Anonymity Mechanisms}\

\noindent\textbf{Randomized Pseudonym Generation.}
To preserve privacy and eliminate persistent identifiers, each SU generates a unique new pseudonym after completing the transmission of its prior pseudonym. These pseudonyms are sampled from a large uniform distribution to ensure randomness and non-linkability. This approach prevents long-term association of pseudonyms with specific users or devices, thus mitigating the risks of adversarial tracking, fingerprinting, or behavioral profiling.

\noindent\textbf{User-Agnostic Logging at PU.}
Rather than identifying individual SUs, the PU focuses exclusively on detecting the presence of watermarked signals. Upon decoding, valid pseudonyms are logged with timestamp and frequency channel. Device-specific metadata are not retained. This design choice protects the anonymity of SUs and simplifies the processing logic of the PU, enabling scalable and low latency operation in high-density deployments.

\subsubsection{False Positive Mitigation}\

\noindent\textbf{Time-Based Expiration Mechanism.}
To maintain the freshness of the database and prevent outdated pseudonyms from persisting, a time-to-live (TTL) policy~\cite{TTLDef} is enforced. Specifically, each pseudonym entry expires $T_0$ seconds after insertion, where $T_0$ is the maximum period allowed by the protocol between SU database queries. This temporal filtering limits the database to recent transmissions, which are most relevant for interference avoidance, and also limits SU query memory usage and search complexity.

\noindent\textbf{Immediate Match-Based Deletion.}
Upon detection of a matching pseudonym during a PU database entry, the older matching entry is deleted. This avoids redundant detection and suppression of a single SU by multiple PUs, helping to reduce query data rates.

\subsection{Pseudonym Detection Framework}
To enable real-time identification of interfering SUs, \Name{} incorporates a pseudonym detection framework at the PU receiver. This framework is designed to to operate under low-SNR, time-varying wireless channels and overlapping SU transmissions.
The detection framework consists of the following steps.
\begin{enumerate}
    \item \textbf{Synchronization}. Incoming complex-valued (IQ) samples are correlated with the data packet preamble to identify the start of a pseudonym frame.
    \item \textbf{Bit Estimation via Pattern Matching}. The average power per chip period is computed and correlated with a predefined codes, and high correlation determines a p-bit value.
    \item \textbf{Pseudonym Frame Assembly}. The system verifies frame integrity by using a fixed m-sequence preamble and a (31, 26) Hamming code error detection and control.
\end{enumerate}

\subsection{Database \& Scalability}

As the number of SUs and PUs increases, the database faces potential scalability challenges. Specifically, high volumes of interference reports can lead to slower query responses, increased latency, and significant storage overhead, thereby impacting overall system performance. We denote the number of SUs as $U_s$, the average number of reports per user as $R$, and the average size of each report as $S_r$. Note that \Name{} is intended to be used in sharing systems in which interference is rare, and thus $R<<1$. The size of the database $D$ can be formulated as $D=U_s \times R \times S_r$. If $R$ and $S_r$ are considered constants or bounded by constants, the growth of the database size $D$ directly depends linearly on the number of SUs $U_s$. Thus, $D$ can also be formulated as $D=O(U_s)$. 

To address these scalability concerns, we adopt an SQLite database implementation~\cite{sqlite2020hipp}, chosen for its lightweight architecture and efficiency in handling frequent read operations with low overhead. Its embedded design eliminates the need for a separate database server, making it suitable for deployment in resource-constrained or distributed environments.
Storage efficiency is further supported by mechanisms that limit unnecessary data retention. In particular, the time-based expiration policy ensures that outdated interference records are automatically purged after a configurable duration, keeping the database size bounded. In parallel, immediate deletion of matched pseudonyms prevents inactive or resolved entries from lingering in the system. These strategies not only reduce the risk of false positives but also help maintain database responsiveness as the number of users and reports scales.
 
Further, \Name{}'s database can be localized for large deployments. We envision multiple (overlapping) database coverage areas, each serving a constrained geographical area. 

In addition to storage-related performance concerns, database request latency is also influenced by the number of concurrent users. As more users simultaneously query the database, contention and resource sharing can increase response times. This relationship can be captured by the following expression: $T_{resp}=T_b \times (1 + \alpha(U_Q - 1))$, where $T_b$ represents the baseline response time for a single user, $\alpha$ is a scaling factor that quantifies the impact of each additional concurrent user, and $U_Q$ denotes the number of users querying concurrently. In the worst case, this relationship simplifies to $T_{resp}=O(U_Q)$, indicating that response time grows linearly with the number of concurrent queries.
Our evaluation indicates that SQLite handles concurrent read operations efficiently, with relatively low overhead even under high query loads. In contrast, concurrent writes introduce more significant performance penalties. However, this asymmetry aligns well with our usage model: SUs frequently read from the database to check for interference, whereas PUs write to the database only when interference is detected, a comparatively infrequent event.
Moreover, recent studies~\cite{gaffney2022sqlite} have shown that SQLite demonstrates robust performance under concurrent access patterns and, in many cases, outperforms alternative systems such as DuckDB. This further validates our choice of SQLite as a scalable and reliable solution for supporting the system’s query-intensive workload.

\subsection{Security and Privacy}

One of the key privacy protections in the system is the use of a pseudonym, which is a random bit string unrelated to any identifiable information about the transmitter. This design ensures that the pseudonym itself does not reveal the identity of the transmitting device. While the data signal inherently leaks some information to nearby eavesdroppers due to its physical characteristics, the pseudonym --- being randomly generated and distributed over a broader area --- does not amplify this risk. It provides a layer of anonymity that protects the transmitter’s identity from passive observers.

However, privacy threats still exist. An eavesdropper may detect a transmission from a SU and successfully demodulate the pseudonym embedded in the watermarked signal. Although obtaining the pseudonym alone does not directly compromise privacy, it enables more sophisticated attacks. For instance, a man-in-the-middle (MitM)~\cite{MIM-UMTS} attacker could relay the captured pseudonym to a malicious transmitter located near the PU. This adversarial transmitter could then re-transmit using the same pseudonym, causing it to be recorded in the interference report database. If the SU assumes the interference report in the database corresponds to its own transmission, \Name{} might unjustly block the non-interfering SU from future access to that channel.

This MitM attack functions similarly to a jamming attack: both disrupt channel access and force legitimate transmitters to vacate the band. From a defensive perspective, the system can mitigate MitM attacks by leveraging the timestamps in interference reports. If the re-transmitted pseudonym is delayed beyond the expected OTA propagation time, it can be flagged as suspicious and discarded. Moreover, like jammers, MitM attack devices are actively transmitting and can therefore be identified through source localization techniques. Future work could enhance system resilience by providing operators with tools to locate and neutralize such adversarial devices.

Beyond over-the-air threats, the database itself presents a potential attack surface. An adversary might attempt to corrupt the database by inserting fake pseudonyms or by disabling it entirely. Inserting bogus pseudonyms would have a similar effect to the MitM attack, polluting the interference record and triggering inappropriate access restrictions. 
To protect the integrity of the database, the system employs two key mechanisms: role-based access control (RBAC) and secure, authenticated APIs. RBAC ensures that only designated PUs, or their trusted proxies, are authorized to write to the database. All other parties, including SUs and external entities, are restricted to read-only access or are entirely denied access. Secure APIs complement this by enforcing strict authentication, input validation, and logging of write operations, preventing tampering by unauthorized actors and enabling traceability of each update.
In contrast, a denial-of-service (DoS) attack on the database could prevent its operation entirely. To improve robustness against such attacks, deploying redundant database replicas or distributed storage mechanisms may offer effective protection.


\section{Implementation}
\label{implementation}

We implemented the full \Name{} system with the purpose of quantifying its performance on our design goals for the system. We implemented \Name{} on the POWDER wireless testbed~\cite{Breen2020powder, powder} because of its large scale deployment, and available software-defined radios (SDRs). In this section, we describe the experimental setup, the SDR implementation, the PU, the SU, and the interference report database. We implement other watermarking schemes for purposes of comparison, and to investigate how the performance is related to our design choices. Our deployment experiment and code are publicly available~\cite{stopsec2025}, and can easily be re-run. 

\subsection{Experimental Setup}
\begin{figure}[tbhp]
  \centering
  \includegraphics[width=0.8\linewidth]{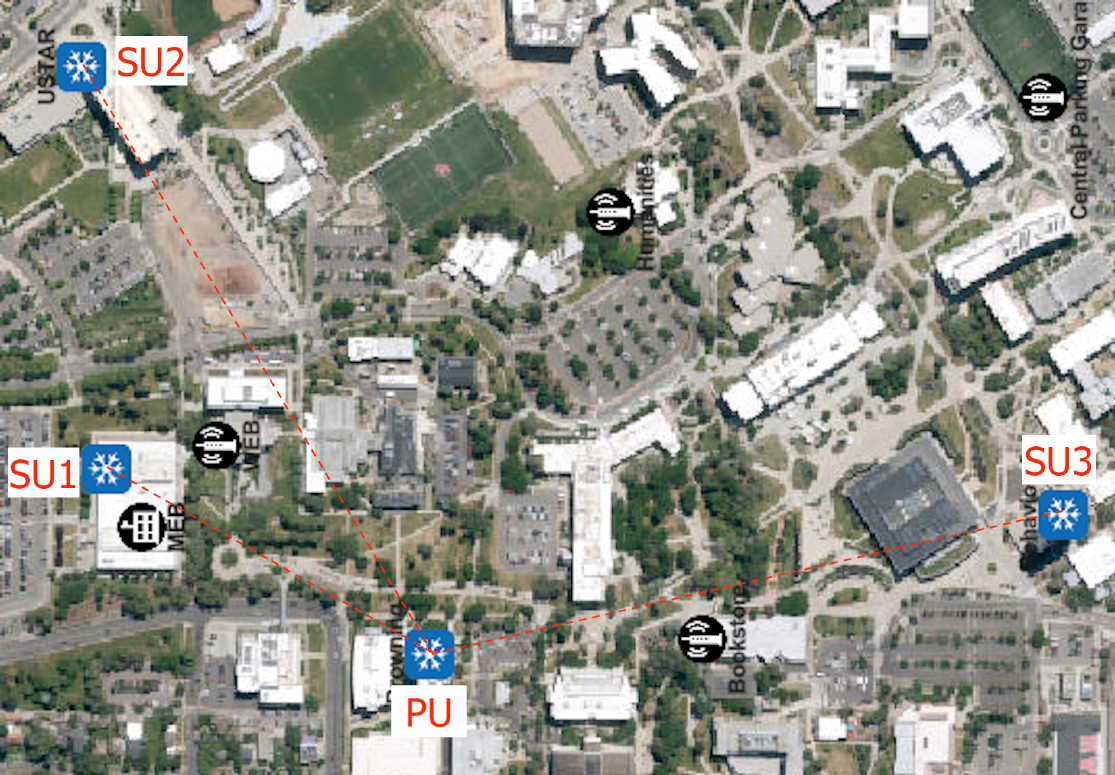}
  \caption{Experiment map: SUs (SU1, SU2 and SU3) all interfere with the PU (bottom middle).}
  \label{fig:Exp.setup}
\end{figure}

\noindent We deployed our experimental setup on the POWDER wireless testbed using four rooftop base stations. Each base station is equipped with a National Instruments (NI) USRP X310 SDR and an associated compute node for transmission and reception processing. Three of the base stations serve as SUs, while the fourth functions as the PU. As illustrated in  Figure~\ref{fig:Exp.setup}, SU1, SU2, and SU3 are located approximately 330m, 590m and 570m from the PU, respectively. All transmissions operate at a center frequency of 3.385\,GHz within the lower part of the cellular C-band. Depending on the experiment, SUs transmit OFDM waveforms on channel bandwidths ranging from 1\,MHz to 10\,MHz,  spanning from 64 to 256 subcarriers.
All system components --- including watermark embedding, pseudonym detection, interference database access, and suppress --- control were developed in Python using UHD APIs for low-level radio control and signal processing. This setup enables end-to-end evaluation of \Name{}'s performance in real-world, outdoor deployment. 

\subsection{Pseudonym Watermarking Implementation}
A key challenge in \Name{} is how to embed pseudonyms into SU transmissions in a way that is both robust to low-SNR conditions and minimally disruptive to data demodulation on the secondary link. This is particularly difficult in time-varying wireless channels, where fading and noise can impair watermark recovery at the PU. To explore the trade-offs between detection reliability and data integrity, we implement two watermarking schemes: (i) a full-band CM approach that distributes watermark energy across all subcarriers, and (ii) a single-subcarrier CM scheme which is what we refer to as the \Name{} approach, which isolates pseudonym transmission to a dedicated subcarrier. This allows us to experimentally quantify the benefits and drawbacks of both strategies that we propose in \Name{}, i.e., coded modulation, and single subcarrier. 

\begin{figure}[tbhp]
  \centering \includegraphics[width=0.9\linewidth]{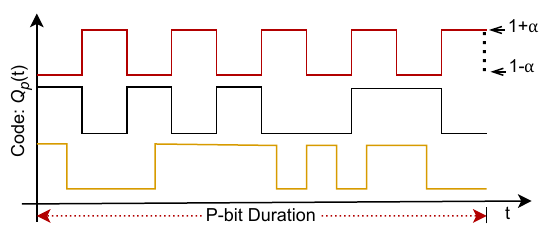}
  \caption{Example chip-level code patterns in \Name{}. Top: an alternating chip amplitude pattern with a chip length of 10. Middle: a truncated m-sequence, chip length 10. Bottom: a full-length m-sequence, chip length 15.}
  \label{fig:codes}
\end{figure}

The code patterns shown in Figure~\ref{fig:codes} are representative variants generated using the general watermarking construction in (\ref{E:watermark}). 
We selected these particular instances to test different code implementations with sufficient chip-to-chip transitions, which aid in removal of the temporal channel variations, and with similar number of low and high values. 

 Each SU transmits one pseudonym bit per data packet. Each SU data packet consists of 100 OFDM symbols, with the number of subcarriers varying based on the transmission bandwidth. For example, in the 2\,MHz configuration, we use 64 OFDM subcarriers: 48 for data, 4 for pilots, and 12 as guard bands. All data subcarriers are modulated using Quadrature Phase Shift Keying (QPSK).
For OFDM packet synchronization, the OFDM packet starts with the high-throughput short training field (HTSTF) symbols, as defined in the 802.11n standard~\cite{ieee80211n}. The HTSTF symbols are not watermarked to maintain signal integrity for OFDM packet synchronization. The watermarked OFDM samples are then sent into the X310 and transmitted at a 3.385\,GHz carrier frequency. We use an RF bandwidth of 1\,MHz to 10\,MHz  with subcarrier spacing of 15.625\,kHz to 39.06\,kHz. 

\subsection{Pseudonym Reception and Reporting}
\begin{figure*}[t]
  \centering
  \includegraphics[width=0.8\textwidth]{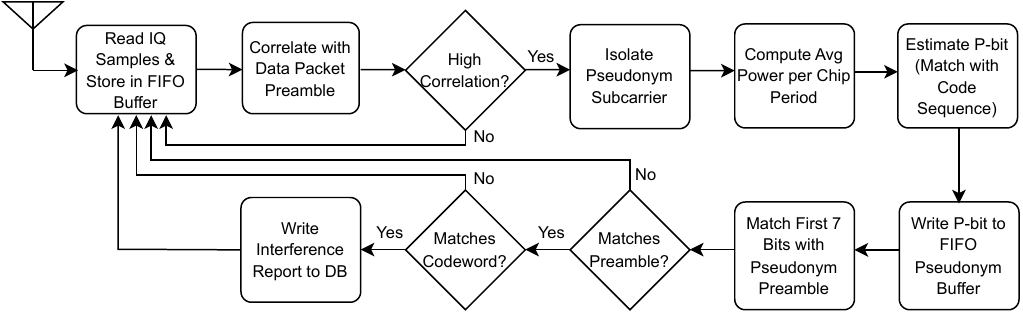}
  \caption{Pseudonym detection and reporting flowchart. The PU processes IQ samples to detect interference, demodulate and decode pseudonym bits, and log interference reports in real time.}
  \label{fig:Exp.flowchart-detection}
\end{figure*}

To enable testing of the PU, we implement a real-time signal processing framework that detects interference, demodulates the watermarked pseudonym bits from the interfering SU signals, processes the pseudonym frame, and reports it to a remote database for enforcement. Figure~\ref{fig:Exp.flowchart-detection} provides a simplified overview of this process. 
We implemented the pipeline entirely in Python, interfaced with USRP X310 devices via the UHD API. These implementation choices are made for ease of use by others, rather than for speed optimization. This implementation runs in real time on a Dell R430 server equipped with Two Intel E5-2630v3 8-Core CPUs at 2.4\,GHz, and serves as the core processing logic at the primary receiver.

\subsubsection{Interference Detection via Adaptive Thresholding.} We implement an adaptive thresholding approach to distinguish watermarked signals from noise. Our method is similar to those used in constant false alarm rate (CFAR) detection in radar and cognitive radio systems~\cite{wikiCFAR,salahdine2016matched}. The thresholding technique used here is adaptive peak-to-median ratio thresholding with dynamic tuning based on runtime feedback. We declare detection if $\textit{peak\_correlation} \geq \textit{threshold\_factor} \times \textit{median\_correlation}$. The threshold factor starts at $4.5$ --- an empirical value--- and adapts dynamically based on detection history. This adaptive thresholding approach is very critical in \Name{} which is designed to operate under varying noise floors and low SNR environments --- helping it maintain robust detection performance even under varying channel conditions.  

The process begins with collecting IQ samples from an USRP X310 receiver. These samples are first buffered in the USRP's onboard memory and then streamed sequentially into a sliding FIFO buffer for real-time processing. The system continuously correlates the contents of the FIFO buffer with a known HTSF preamble. If the threshold condition is not satisfied, new samples are fetched from the USRP and the buffer is updated. The PU receiver remains in a continuous listening mode, fetching and evaluating samples in real time. When the correlation score exceeds the adaptive threshold, it indicates the likely presence of an interfering signal containing a pseudonym watermark, prompting the system to initiate further decoding and identification steps.

\subsubsection{Interference Identification.} Upon detecting a strong correlation, the system isolates the pseudonym subcarrier using an FFT operation. The absolute magnitude values of six samples on the pseudonym subcarrier, extracted from six OFDM symbols, are averaged and pushed into a per-chip FIFO buffer. Once the FIFO buffer is filled, it is correlated with the code sequence used for code modulation to decode each P-bit. The decoded P-bit is appended to a pseudonym bit FIFO buffer and the process continues until the buffer is full. The reconstructed frame undergoes validation via a 7-bit m-sequence preamble check. If the preamble is correct, the remaining bits are decoded using a \((31,26)\) Hamming decoder to correct single-bit errors and detect double-bit errors. Only validated pseudonym packets are reported to the database.

\subsubsection{Interference Reporting.} Once a valid pseudonym is decoded, a timestamp is appended, and the interference report is written to a remote database. This database serves as a feedback channel to SUs. If an SU finds its own pseudonym in the database, it infers that it has caused interference and halts transmission on that channel. Note that in our implementation, each interference report includes the decoded pseudonym along with its corresponding timestamp. Since we use a single PU and a single frequency band, incorporating location and frequency information would be redundant.

\subsection{System Configurations for Evaluation}
To assess \Name{}'s performance under diverse operating conditions, we varied key PHY-layer parameters --- including subcarrier counts, bandwidths, and the number of interfering SUs --- as summarized in Table~\ref{tab:config-summary}.

\subsubsection{Testing Across Multiple Subcarrier Settings}

To evaluate \Name{}’s sensitivity to subcarrier allocation, we implement three configurations under a fixed 2\,MHz bandwidth: (i) a single pseudonym subcarrier (31.25\,kHz), (ii) two subcarriers (62.5\,kHz), and (iii) three subcarriers (93.75\,kHz). 
When we use more than one subcarrier in \Name{}, we simply apply the single-subcarrier coded modulation to each subcarrier in the OFDM packet identically.
All other transmission settings remain fixed. This setup helps assess the trade-off between detection robustness and spectrum overhead.

\subsubsection{Transmission Bandwidth and Subcarrier Configuration}
\Name{} can be applied to OFDM signals of any bandwidth. We tested \Name{} under three bandwidth settings: 2\,MHz, 5\,MHz and 10\,MHz, corresponding to 64, 128, and 256 OFDM subcarriers, respectively. In all three configurations, a single subcarrier is dedicated to pseudonym watermarking, and the remaining subcarriers are allocated for data, pilots, and guard bands.
The choice of bandwidth affects the subcarrier spacing, which is a critical factor for both watermark robustness and data performance. For example, at 2\,MHz with 64 subcarriers, the subcarrier spacing is approximately 31.25\,kHz, while at 5\,MHz and 128 subcarriers, the spacing increases to 39.06\,kHz. Wider subcarrier spacing improves frequency-domain separation, enhancing watermark detection at low SNR, though it may increase sensitivity to timing offsets at higher data rates.

\begin{table}[t]
  \caption{System Configurations in \Name{} Evaluation.}
  \label{tab:config-summary}
  \centering
  \resizebox{\columnwidth}{!}{%
  \begin{tabular}{@{}ccccc@{}}
    \toprule
    \textbf{Bandwidth} & \textbf{Subcarriers} & \textbf{Subcarrier Spacing} & \textbf{Pseudonym Bandwidth} & \textbf{Concurrent SUs} \\
    \midrule
    2 MHz  & 64  & 31.25 kHz & 1 / 64 = 1.5\%  & 1--3 \\
    5 MHz  & 128 & 39.06 kHz & 1 / 128 = 0.8\% & 1--3 \\
    10 MHz & 256 & 39.06 kHz & 1 / 256 = 0.4\% & 1--3 \\
    \bottomrule
  \end{tabular}
  }
\end{table}

\subsubsection{Scaling to Multiple Interfering SUs}
In the real world, sometimes, a second (or third) SU will start interfering with a PU before the system can shut off the first interfering SU. \Name{} must be able to handle this case. To evaluate this scalability, we test \Name{} with one, two, and three simultaneously active and interfering SUs. We use the experiment setup shown in Figure~\ref{fig:Exp.setup}. Each SU transmits with a unique pseudonym but shares the same OFDM configuration. This setup enables us to analyze decoding latency, and detection reliability under increasing interference complexity. 


\section{Evaluation}
\label{Evaluation}
In this section, we evaluate the performance of \Name{} through a series of experiments. Our goal is to quantify how effectively the system can detect, identify, and suppress interfering SUs in real time under various operating conditions.

We begin by analyzing our underlying watermarking schemes --- CM and \Name{} --- on their ability to embed and decode pseudonym bits at the primary receiver under low SNR conditions. As part of this analysis, we compare CM against a baseline PAM watermarking approach to quantify improvements in detection reliability. We also evaluate \Name{} in terms of both its pseudonym detection capability and its impact on data demodulation in the secondary link. 
We then present results demonstrating \Name{}’s responsiveness, measured as the latency between the onset of interference and successful suppression of the 
interfering SU. We evaluate this latency across different watermark bandwidth allocations and SNR levels, highlighting the trade-off between robustness and overhead.
Finally, we assess how the system responds when multiple SUs concurrently interfere with the PU. We analyze both the increase in latency and the probability of successful pseudonym detection under these conditions.

Together, these experiments validate that \Name{} enables real-time identification and suppression of an interfering device, even in applications with very low SNR and multiple interfering users --- making it an effective spectrum sharing protocol for applications that require quick and automatic mitigation of radio frequency interference (RFI).

\subsection{Performance of Watermarking Schemes}
\subsubsection{Evaluation of Code Modulation}
Here we compare the pseudonym detection capabilities of CM and  PAM watermarking schemes. We compare the pseudonyms received in error for the CM modulation described in \ref{subsec:pattern} and the PAM watermarking scheme. In both cases we transmit watermarked OFDM signals using the setup in Figure~\ref{fig:Exp.setup}. At the primary receiver, we compute the probability of pseudonym bit error for each $\EbnoSmall$. The number of pseudonym bits needed to experimentally estimate the probability of bit error is a function of the error probability estimation and confidence level~\cite{Dragan2012calculating}. Here, we use a theoretical P-bit error rate~\cite{Meles2024pseudonymetry} and a confidence level of $99\%$ to determine the minimum number of P-bits needed at each $\EbnoSmall$. 

\begin{figure}[tb]
  \centering
  \includegraphics[width=0.9\linewidth]{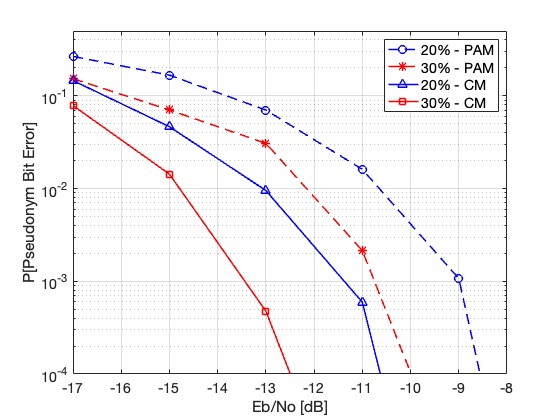}
  \caption{Probability of pseudonym bit error vs.\ $\EbnoSmall$ at the PU receiver, comparing CM and PAM watermarking schemes.}
  \label{fig:pseudonym_ber_pattern}
\end{figure}

Figure~\ref{fig:pseudonym_ber_pattern} shows that the CM watermarking outperforms the PAM watermarking by more than $2\,\mathrm{dB}$. For example, at $\EbnoSmall = -11\,\mathrm{dB}$, the probability of pseudonym bit error for CM is $7\times{10}^{-4}$. For PAM watermarking, a similar probability of pseudonym bit error is achieved at $\EbnoSmall = -9\, \mathrm{dB}$. 

\begin{figure}[tbp]
    \centering
    \includegraphics[width=0.9\linewidth]{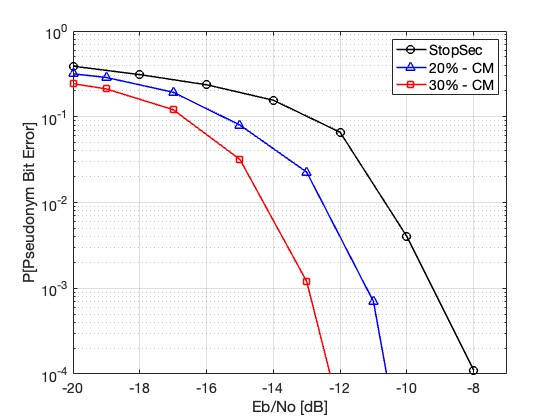}
    \caption{Probability of pseudonym bit error vs.\ $\EbnoSmall$ at the PU. Plots compare CM (20\% and 30\% modulation) vs.\ \Name{}.}
    \label{fig:pseudonym_ber}
\end{figure}

We also compare pseudonym detection capabilities for CM and \Name{}. In both methods, we transmit one pseudonym bit per OFDM packet. Because pseudonym bits are transmitted in only one of the 64 subcarriers, \Name{} has a factor of 64 smaller bandwidth than CM. 
Lower bandwidth is generally able to carry a lower data rate for a given SNR. 
However, because the subcarrier is dedicated to pseudonym transmission, we use a larger modulation index for watermarking, which compensates in part for the small bandwidth in \Name{}.  Figure~\ref{fig:pseudonym_ber} compares pseudonym detection performance for \Name{} and CM. 
Despite a factor of 64 larger bandwidth of watermarking, the CM method results in only a $2\,\mathrm{dB}$ advantage in pseudonym detection performance for 20\% CM, or $4\,\mathrm{dB}$ advantage for 30\% CM. This additional performance for CM comes at the cost of $2-3\,\mathrm{dB}$ degradation in SU communication performance, as we describe next.

\subsubsection{Impact of Watermarking on Data Demodulation}
To assess the impact of watermarking schemes on the performance of SUs, we evaluate the data bit error rate (BER) under three transmission configurations: (i) unwatermarked OFDM (ii) \Name{} watermarking (iii) 20\% CM watermarking. This experiment quantifies the effect of pseudonym embedding on data demodulation accuracy on the secondary link. 
In \Name{}, we watermark only the dedicated pseudonym subcarrier. We use $48$ data subcarriers, $4$ pilot subcarriers and $11$ guard subcarriers. Here, one of the guard bands is used as a pseudonym subcarrier. We employ modulation index of $m = 1.0$ on the pseudonym subcarrier. 
For each configuration, we transmit OFDM packets over the secondary link and collect raw IQ samples at the receiver. The receiver performs standard OFDM baseband processing: symbol synchronization, channel equalization, 64-point FFT, and QPSK demapping to recover transmitted data bits. The experiment is repeated across a range of $\EbnoSmall$ values. In this section, we use $\EbnoSmall$ to indicate the energy per data bit. To estimate $\EbnoSmall$, we measure noise power by sampling the channel when the transmitter is idle, and compute signal power from received packets during transmission. BER is computed by comparing received bits against a known reference and calculating the ratio of bit errors to total bits received. To determine the number of data bits to send in order to reliably estimate the BER, we use an estimation method described in this work~\cite{Dragan2012calculating}.

\begin{figure}[tbp]
  \centering
  \includegraphics[width=0.9\linewidth]{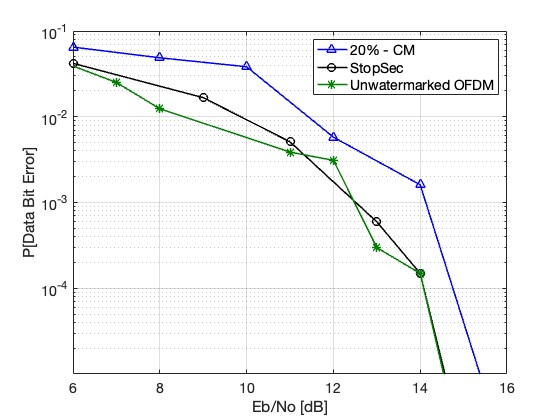}
  \caption{OFDM receiver probability of data bit error vs.\ $\EbnoSmall$ at the SU data receiver. Compares CM (20\% modulation index), \Name{}, vs.\ unwatermarked signal.}
  \label{fig:data_ber}
\end{figure}

Experimental results shown in Figure~\ref{fig:data_ber} demonstrate that \Name{} does not degrade data demodulation at the secondary receiver whereas 20\% CM degrades performance by up to $2\,\mathrm{dB}$. This validates that \Name{} does not affect data demodulation at the secondary receiver. However, as indicated in Figure~\ref{fig:pseudonym_ber}, this is at the cost of higher probability of pseudonym bit error at the primary receiver. 

\subsection{End-to-End Latency Evaluation}
We evaluate the total latency from the onset of interference to the point at which the interfering SU is successfully stopped. This measurement captures the complete system response time, including: Interference detection at the PU, pseudonym decoding from the watermarked subcarrier, write and lookup operations in the remote interference database, and delay until the SU query and stoppage of use of the band.  Timestamps are recorded at each stage of the pipeline, enabling us to compute the total detection-to-suppression delay. This metric characterizes the responsiveness of the \Name{} system under realistic deployment conditions.

\begin{figure}[tbp]
  \centering \includegraphics[width=0.9\linewidth]{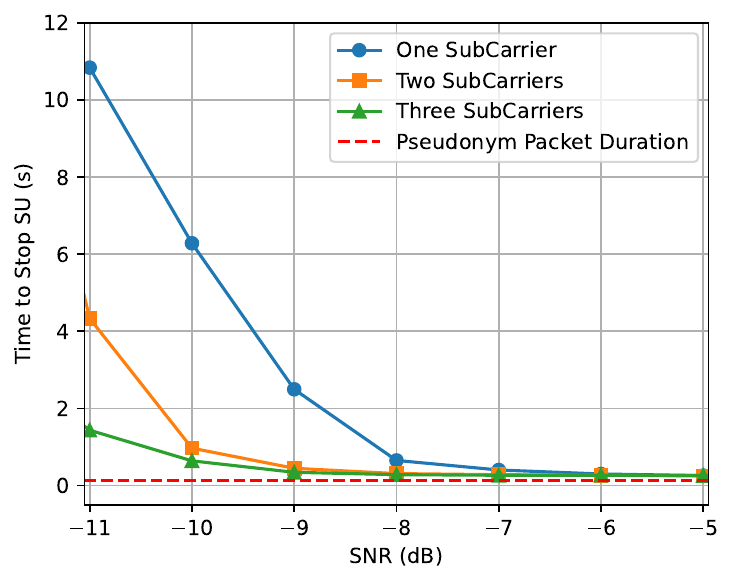}
  \caption{Stopping latency vs.\ SNR. When \Name{} can detect interfering SUs at lower SNR by using more (2-3) subcarriers.}
  \label{fig:latency-subcarrier}
\end{figure}

Using the experimental setup illustrated in Figure~\ref{fig:Exp.setup}, we evaluate the latency of the \Name{} system.  
In this experiment, SU1 acts as the interfering secondary user, while PU serves as the primary user monitoring the channel. 
We measure this latency across different watermarking subcarrier allocations and SNR levels. Figure~\ref{fig:latency-subcarrier} presents the results. The x-axis represents the SNR (in $\mathrm{dB}$), and the y-axis shows the latency (in seconds), i.e., the time it takes for \Name{} to stop the SU from transmitting in the band. 

As shown in Figure~\ref{fig:latency-subcarrier}, the single-subcarrier configuration achieves an average latency of less than 270 milliseconds when the SNR exceeds $-8\,\mathrm{dB}$. In contrast, both the two-subcarrier and three-subcarrier configurations maintain similar latency even at lower SNR levels ($-10\,\mathrm{dB}$ and above), demonstrating an improvement in performance of approximately ($2\,\mathrm{dB}$). These results indicate first that a single subcarrier \Name{} system can reliably detect and stop interference caused by a SU within a few hundred ms, even when the interference is 8 dB below the noise power.  Further, the results show how \Name{} could be adapted to even lower SNR scenarios by adding additional subcarriers --- 2-3 subcarriers allow detection with 2 dB less SNR. 

\subsection{Impact of SU Transmission Bandwidth}

To be incorporated into future spectrum sharing systems, \Name{} must work with secondary systems that operate across a wide range of RF bandwidths. We contend that  pseudonym detection and SU stopping performance is primarily determined by the pseudonym subcarrier bandwidth. To test this, we evaluated \Name{} under three different transmission bandwidths - 2\,MHz, 5\,MHz, and 10\,MHz - using 64, 128, and 256 subcarriers, respectively, in order to keep the subcarrier bandwidth relatively constant. In all configurations, a single subcarrier is allocated for pseudonym transmission, thus the pseudonym bandwidth overhead is approximately 1.5\% (2\,MHz), 0.7\% (5\,MHz), and 0.4\% (10\,MHz).

\begin{figure}[tbp]
  \centering
  \includegraphics[width=0.9\linewidth]{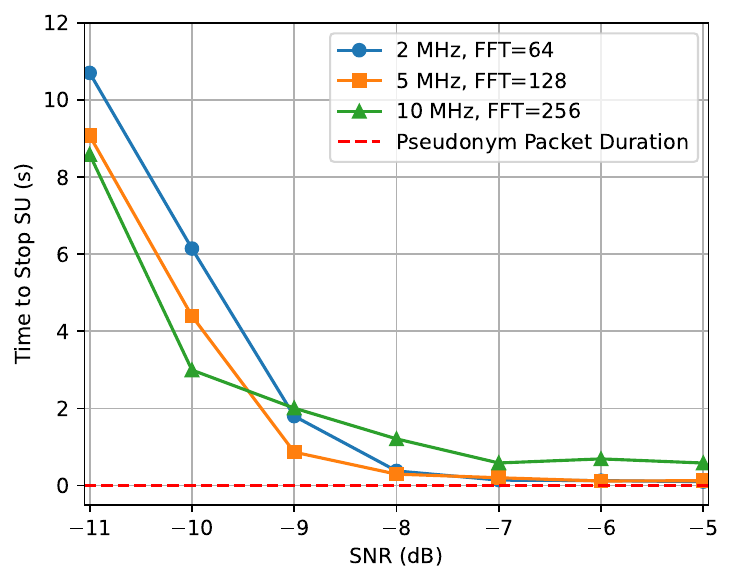}
  \caption{Latency vs.\ SU signal bandwidth and number of OFDM subcarriers.}
  \label{fig:latency-bw}
\end{figure}

As shown in Figure~\ref{fig:latency-bw}, \Name{} maintains robust interference detection and suppression performance across all bandwidth settings. Notably, the 5\,MHz configuration demonstrates slightly superior performance compared to the 2\,MHz case. This can be attributed to the somewhat larger subcarrier spacing in 5\,MHz (39.06\,kHz) compared to the 31.25\,kHz subcarrier spacing in the 2\,MHz setting.

However, in the 10\,MHz configuration, we observe an increase in latency under high SNR conditions (above $-9\,\mathrm{dB}$). This degradation is due to the larger sampling rate required for processing 10\,MHz transmissions, which increases the data throughput and computational complexity. At some rate, our real-time implementation of \Name{} drops some samples, and thus pseudonym packets, before detection. This performance bottleneck is likely not a fundamental limitation, but an artifact of the low-compute implementation. It can be mitigated by deploying a more efficient detection algorithm or increasing processing resources at the primary receiver.

The results indicate that \Name{} can be used across a range of SU signal bandwidths. Further, the required watermarking overhead decreases as a percentage of transmitted signal as the bandwidth increases, but \Name{} performance remains approximately constant.

\subsection{Stopping Multiple Interfering SUs}
\subsubsection{Evaluation of Latency Under Varying Number of Interfering SUs}
\begin{figure}[tbp]
  \centering
  \includegraphics[width=0.9\linewidth]{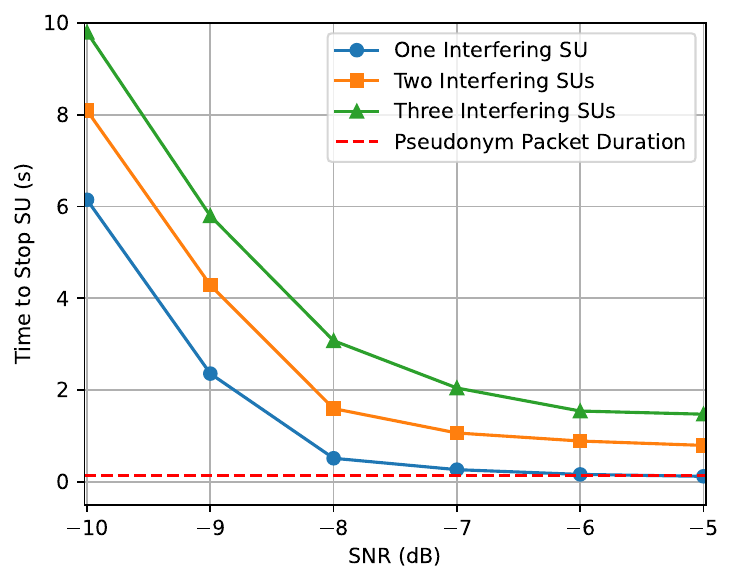}
  \caption{Latency to stop all SUs vs.\ SNR and number of interfering SUs.}
  \label{fig:latency-RFI}
\end{figure}

In real-world deployments, multiple SUs may interfere with a PU before the first is shut down. To evaluate \Name{}’s scalability, we test scenarios with one, two, and three concurrently active SUs. We use the experiment setup shown in Figure~\ref{fig:Exp.setup}. Each SU randomly generates pseudonym packet, but shares the same OFDM settings. We measure the latency at each SNR for three scenarios: (i) When only SU1 is transmitting (ii) when SU1 and SU2 are concurrently transmitting (3) when all three are simultaneously transmitting. The PU runs the same detection algorithm for all three scenarios --- listen to the channel, decode pseudonym when interference is detected, and write interference report to the database. 
Figure~\ref{fig:latency-RFI} shows the latency evaluation result. The y-axis represents the latency in seconds, while the x-axis denotes the SNR at the PU in $\mathrm{dB}$.

As expected, \Name{} exhibit increased latency in the scenarios with more interfering SUs due to two reasons: (i) pseudonym decoding is difficult when there are multiple interfering signals (2) the PU stops one SU and then the next until all SUs are shut off. The system achieves the lowest latency in the single-SU case, followed by two-SU and three-SU scenarios, respectively. Note that in all three scenarios, latency is measured from the first arrival of interference at the PU to the time all interfering SUs are successfully stopped. Across all cases, however, latency remains below ten seconds when the SNR is greater than or equal to $-10\,\mathrm{dB}$. This demonstrates the system’s robustness and real-time responsiveness even under moderately to highly degraded channel conditions.
During the experiment, SU1 --- whose signal is strongest at the PU --- was detected and mitigated first, followed by SU2 and then SU3. Notably, although SU3 is physically closer to the PU than SU2, it was stopped last due to the higher channel loss in its channel to the PU. This confirms that our system prioritizes interference mitigation based on signal impact rather than proximity alone.

\subsubsection{Probability of Pseudonym Packet Detection}
\begin{figure}[tbp]
  \centering
  \includegraphics[width=0.9\linewidth]{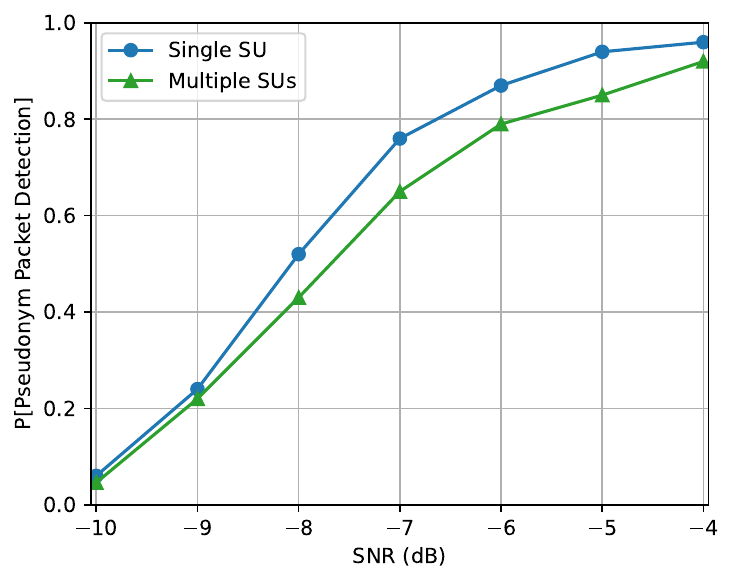}
  \caption{Probability of pseudonym packet detection vs.\ SNR, comparing one interfering SU and 3 interfering SUs.}
  \label{fig:detection-RFI}
\end{figure}

Figure~\ref{fig:detection-RFI} presents the probability of successful pseudonym packet detection under two different interference scenarios: a single interfering SU and three concurrent interfering SUs. The y-axis denotes the probability of detection, while the x-axis represents the SNR at the PU in dB. As expected, the detection probability is consistently higher in the single-SU scenario compared to the three-SU scenario. This is due to increased interference in the presence of multiple concurrent transmissions, which degrades the PU’s ability to reliably decode pseudonym packets.

In our context, a pseudonym packet is considered successfully detected only if it is transmitted by a SU, correctly decoded by the PU (including detection of the preamble and passing of the forward error correction check), logged in the remote database, and subsequently queried and matched with the local database, enabling timely suppression of the interfering SU. The observed degradation in detection probability under multiple interfering SUs highlights that it would likely take somewhat more time to suppress an interfering SU when multiple SUs are simultaneously interfering.

\subsection{Database Scalability Evaluation}

\begin{figure}[tbp]
  \centering
  \includegraphics[width=0.9\linewidth]{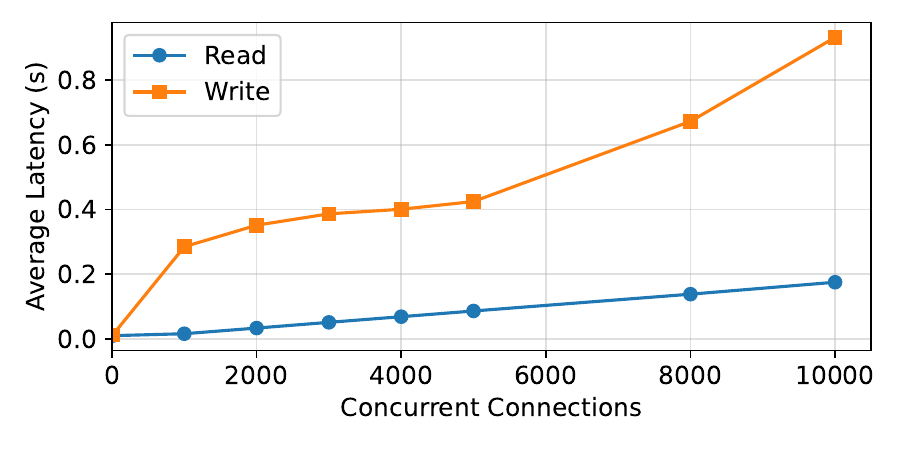}
  \caption{Avg.\ database response time vs.\ concurrent connections.}
  \label{fig:database_eval}
\end{figure}

To assess the scalability of the database, we conduct experiments using Apache Bench~\cite{apachebench}, a widely used benchmarking tool that measures how many requests a server can handle per second. Specifically, we evaluate the database’s response time under varying levels of concurrency to understand its behavior under load.
Figure~\ref{fig:database_eval} presents the response time overhead for both read and write operations across different levels of concurrency. As expected, concurrent writes exhibit significantly higher latency compared to concurrent reads. This is due to SQLite’s architectural constraint of allowing only a single writer at a time, which forces write operations to be serialized. Consequently, as the number of concurrent writers increases, contention grows and delays accumulate.
In contrast, concurrent read operations scale much more efficiently. Even under extreme conditions, such as $10^4$ simultaneous read connections, the average response time is under 200 ms. This behavior is particularly favorable in our deployment context: write operations occur only sporadically, when a PU detects interference, whereas SUs continuously query the database to monitor channel availability.
As a result, the read-dominant workload pattern in our system aligns well with SQLite’s strengths. Despite the inherent limitations on write concurrency, the overall database performance remains robust and does not become a bottleneck for \Name{}, even as the number of SUs scales substantially.

\section{Related Work}
\label{sec:Related Work}
\noindent Maximizing the efficiency of radio spectrum utilization is a priority for spectrum policy, science, and the wireless industry. Towards this goal, there have been substantial developments in wireless technology, spectrum regulation and policy that increase the utility of the spectrum. In this section, we present a summary of existing spectrum sharing strategies and their RFI mitigation techniques.

\noindent \textbf{RFI Detection and Suppression Schemes.}
One of the oldest ways to mitigate interference is through RFI detection and suppression. This method is also referred to as a ``unilateral'' approach~\cite{committee2010spectrum} since there is no coordination between the transmitter and the receiver in the mitigation process. In this technique, the receivers use filtering, excision and cancellation techniques to reduce the effect of RFI on the composite signal~\cite{Barnabaum1998new,Leshem1999detection,Mahmoud2017nonlinear,Mahmoud2018impact}. 
These unilateral approaches do not require participation of the transmitters to mitigate RFI, but they cannot be effective when spectrum occupancy is high and the nature of the RFI sources are unknown. Even for known sources, RFI suppression is imperfect due to an unknown and time-varying frequency-selective wireless channel, which can not be perfectly estimated, in particular at low SNR. 

\noindent \textbf{Power Control Schemes.}
Power control schemes enable reuse of the spectrum at different geographical locations by reducing transmit powers and selection of wireless system use cases. For example, the International Telecommunication Union (ITU) recommends that active transmitters operate below predefined interference power levels, setting a threshold for the amount of permissible interference received by the licensed protected band users~\cite{ITU-RA769-2,ITU-RS2017-0}. 
Although the power control method continues to be a critical component of interference mitigation and spectrum sharing, it can be too conservative when imperfect channel models are used. Power control schemes can be used in conjunction with other strategies to maximize the use of the radio spectrum and facilitate coexistence among diverse wireless systems.

\noindent \textbf{Cooperative Spectrum Sharing Schemes.}
``The only practical way to satisfy the demands of all applications --- commercial, scientific and federal --- is to encourage spectrum sharing among incumbent users and new entrants''~\cite{FCC2023TAC}. It is critical that the spectrum is dynamically shared and successful sharing  depends on coordinated efforts by all stakeholders --- incumbent users and new entrants. Coordinated spectrum sharing can benefit spectrum users by allowing one user to access other users' bands at suitable times~\cite{committee2010spectrum,Martikainen2023DSS}. However, one big challenge in cooperative spectrum sharing is interference control. Existing spectrum sharing control systems focus primarily on \textit{ex ante} interference prevention~\cite{NSF2024spectrum}. Future interference control systems must include interference mitigation mechanisms to deal with harmful interference if it occurs. \Name{} is a closed-loop cooperative sharing protocols that reacts to interference to PU from SUs --- mitigating interference in real-time.

\noindent \textbf{Automated Frequency Coordination System.}
The AFC system is a cooperative spectrum-sharing framework that allows unlicensed standard-power devices to access the 6\,GHz shared spectrum~\cite{FCC2024AFC,Intel2024AFC}. By creating exclusion zones around PUs based on geolocation and propagation models~\cite{ISED2022Canada}, the AFC protects incumbent services from interference~\cite{park2024AFC}. The AFC is critical for enabling license-exempt services such as Wi-Fi. However, its ``static and conservative'' exclusion zones still result in inefficient spectrum usage~\cite{Aarushi2024RDZ}. Moreover, the AFC has no mechanism to identify or suppress a specific SU transmitter that causes interference to a PU~\cite{marsh2020att}.

\noindent \textbf{Spectrum Access System.}
The spectrum access system (SAS) is a spectrum coordination system that manages the Citizens Broadband Radio Service (CBRS) band of 3550--3700\,MHz~\cite{Google2024SAS} in the US. It is a cloud based service that manages the wireless communications of devices transmitting in the CBRS band, across multiple tiers of priority~\cite{Sohul2015SAS}. When an incumbent is detected, the SAS removes or reassigns devices from the affected frequency band and geographic area to prevent harmful interference. This process leverages both sensing --- via the Environmental Sensing Capability (ESC) network --- and propagation modeling. For managing coexistence between Priority Access License (PAL) and General Authorized Access (GAA) users, the SAS relies on device registration data and propagation models to assign channels and enforce geographic separation~\cite{WInnForum_TS-0016, WInnForum_CBRSTechReq}.
Band rules mandate the deployment of an ESC network, with each sensor surrounded by an exclusion zone. The SAS enforces these zones by creating model-based protections around higher-priority users when they are active on a channel. However, this approach does not guarantee complete interference prevention, as it does not identify or suppress specific interfering devices.

\noindent \textbf{Reactive Interference Control.}
Some spectrum sharing mechanisms rely on static databases or pre-allocated frequency use models to prevent interference between types of users. Others use reactive interference control, in which systems respond to real-time observations of interference. Reactive approaches have emerged as a critical means for efficient sharing in environments with unpredictable or intermittent interference.
For example, the reactive framework in~\cite{Aarushi2024RDZ} determines when transmitters in a Radio Dynamic Zone (RDZ) should be turned off, driven by spectrum measurements conducted near protected users. It shares our goal of limiting transmission based on actual interference rather than predicted or modeled interference, but it lacks a mechanism to identify the actual interferer(s), and instead turns off a subset of transmitting devices until the interference is reduced. Our system complements this direction by offering per-transmitter accountability and control, making it a promising enabler for next-generation reactive spectrum governance. \Name{} enables real-time detection, identification, and targeted suppression of the interfering SU(s). It leverages passive sensing at the PU, RF watermarking, and a distributed database to support precise and scalable interference accountability --- even under low SNR and overlapping transmission scenarios. Compared to CBRS, \Name{} achieves finer granularity and does not assume static location information or persistent device identifiers, making it more suitable for dynamic and heterogeneous wireless environments.


\section{Conclusion and Future Work}
\label{sec:discussion}

\Name{} provides a new practical and scalable capability to build dynamic spectrum sharing systems that allow PUs to detect and stop interfering SUs. This capability is highly efficient for the SUs, as it occupies a very low percentage of their OFDM communication signal bandwidth, e.g., 0.4\% for a 10 MHz channel. The watermark does not degrade the performance of the SU communication link like prior methods. The proposed watermark can be detected at the PU in real-world over-the-air channels even when the interfering SU signal is 10 dB below the noise floor, as a result of its low rate coded modulation. Multiple simultaneously interfering SUs can be detected and stopped. This capability addresses a key \textit{double deficiency} encountered in sharing system design: over-restricting cooperative SUs while under-protecting PUs. In \Name{}, the sharing protocol does not need vast exclusion zones for SUs, and it automatically protects PUs by removing the interfering SUs from the band. However, \Name{} is precise --- forcing only those SUs that have been found to interfere to change band. 

We implemented and evaluated \Name{} on the POWDER testbed using a single PU, three SUs, and a database server on a 2.4\,GHz CPU PC. While modest, this deployment demonstrates the system's practical feasibility. Importantly, the design is scalable, supporting \Name{} deployment at regional or national levels, and aligning with emerging trends in cloud-based spectrum coordination. \Name{} relies on an internet-connected interference database, an architectural choice increasingly common in modern systems like CBRS’s Spectrum Access System (SAS) and the 6 GHz band's AFC. This positions \Name{} as a technically aligned solution for emerging cloud-integrated spectrum governance frameworks. 
By enabling real-time, SU-specific interference accountability, \Name{} bridges a critical gap between enforcement precision and spectrum efficiency --- paving the way for reliable, accountable and scalable cooperative spectrum sharing.

While \Name{} demonstrates promising results, several directions remain for future work. First, we plan to extend the system to support mobility, where both PUs and SUs are mobile. This requires adapting pseudonym detection to account for fast fading related to mobility and rapidly changing SNR profiles. Second, we aim to explore more advanced pseudonym encoding schemes that can improve robustness without increasing bandwidth overhead, including machine learning-based interference identification and decoding.
Finally, broader-scale testing across different spectrum bands and wireless systems can help validate \Name{}'s adaptability and effectiveness in diverse deployment scenarios.

\bibliographystyle{ACM-Reference-Format}
\bibliography{Ref}


\begin{thebibliography}{46}


\ifx \showCODEN    \undefined \def \showCODEN     #1{\unskip}     \fi
\ifx \showDOI      \undefined \def \showDOI       #1{#1}\fi
\ifx \showISBNx    \undefined \def \showISBNx     #1{\unskip}     \fi
\ifx \showISBNxiii \undefined \def \showISBNxiii  #1{\unskip}     \fi
\ifx \showISSN     \undefined \def \showISSN      #1{\unskip}     \fi
\ifx \showLCCN     \undefined \def \showLCCN      #1{\unskip}     \fi
\ifx \shownote     \undefined \def \shownote      #1{#1}          \fi
\ifx \showarticletitle \undefined \def \showarticletitle #1{#1}   \fi
\ifx \showURL      \undefined \def \showURL       {\relax}        \fi
\providecommand\bibfield[2]{#2}
\providecommand\bibinfo[2]{#2}
\providecommand\natexlab[1]{#1}
\providecommand\showeprint[2][]{arXiv:#2}

\bibitem[wik(2024)]%
        {wikiCFAR}
 \bibinfo{year}{2024}\natexlab{}.
\newblock \bibinfo{title}{Constant false alarm rate}.
\newblock \bibinfo{howpublished}{\url{https://en.wikipedia.org/wiki/Constant_false_alarm_rate}}.
\newblock
\newblock
\shownote{Wikipedia article}.


\bibitem[Abdelgelil and Minn(2018)]%
        {Mahmoud2018impact}
\bibfield{author}{\bibinfo{person}{Mahmoud Abdelgelil} {and} \bibinfo{person}{Hlaing Minn}.} \bibinfo{year}{2018}\natexlab{}.
\newblock \showarticletitle{Impact of Nonlinear {RFI} and Countermeasure for Radio Astronomy Receivers}.
\newblock \bibinfo{journal}{\emph{IEEE Access}}  \bibinfo{volume}{6} (\bibinfo{year}{2018}), \bibinfo{pages}{11424--11438}.
\newblock
\urldef\tempurl%
\url{https://doi.org/10.1109/ACCESS.2018.2808414}
\showDOI{\tempurl}


\bibitem[Abdelgelil and Minn(2017)]%
        {Mahmoud2017nonlinear}
\bibfield{author}{\bibinfo{person}{Mahmoud~E. Abdelgelil} {and} \bibinfo{person}{Hlaing Minn}.} \bibinfo{year}{2017}\natexlab{}.
\newblock \showarticletitle{Non-Linear Interference Cancellation for Radio Astronomy Receivers with Strong {RFI}}. In \bibinfo{booktitle}{\emph{IEEE Global Communications Conf.}} \bibinfo{pages}{1--6}.
\newblock
\urldef\tempurl%
\url{https://doi.org/10.1109/GLOCOM.2017.8254541}
\showDOI{\tempurl}


\bibitem[Barnabaum and Bradley(1998)]%
        {Barnabaum1998new}
\bibfield{author}{\bibinfo{person}{Cecilia Barnabaum} {and} \bibinfo{person}{Richard Bradley}.} \bibinfo{year}{1998}\natexlab{}.
\newblock \showarticletitle{A New Approach to Interference Excision in Radio Astronomy: Real-Time Adaptive Cancellation}.
\newblock \bibinfo{journal}{\emph{The American Astronomical Society}} (\bibinfo{year}{1998}), \bibinfo{pages}{2598--2614}.
\newblock


\bibitem[Breen et~al\mbox{.}(2020)]%
        {Breen2020powder}
\bibfield{author}{\bibinfo{person}{Joe Breen}, \bibinfo{person}{Andrew Buffmire}, \bibinfo{person}{Jonathon Duerig}, \bibinfo{person}{Kevin Dutt}, \bibinfo{person}{Eric Eide}, \bibinfo{person}{Mike Hibler}, \bibinfo{person}{David Johnson}, \bibinfo{person}{Sneha~Kumar Kasera}, \bibinfo{person}{Earl Lewis}, \bibinfo{person}{Dustin Maas}, \bibinfo{person}{Alex Orange}, \bibinfo{person}{Neal Patwari}, \bibinfo{person}{Daniel Reading}, \bibinfo{person}{Robert Ricci}, \bibinfo{person}{David Schurig}, \bibinfo{person}{Leigh~B. Stoller}, \bibinfo{person}{Jacobus Van~der Merwe}, \bibinfo{person}{Kirk Webb}, {and} \bibinfo{person}{Gary Wong}.} \bibinfo{year}{2020}\natexlab{}.
\newblock \showarticletitle{{POWDER}: Platform for Open Wireless Data-driven Experimental Research}. In \bibinfo{booktitle}{\emph{"Proceedings of the 14th International Workshop on Wireless Network Testbeds, Experimental Evaluation and Characterization (WiNTECH)"}}.
\newblock
\urldef\tempurl%
\url{https://doi.org/"10.1145/3411276.3412204"}
\showDOI{\tempurl}


\bibitem[Developers(2025)]%
        {stopsec2025}
\bibfield{author}{\bibinfo{person}{StopSec Developers}.} \bibinfo{year}{2025}\natexlab{}.
\newblock \bibinfo{title}{StopSec System}.
\newblock \bibinfo{howpublished}{\url{https://github.com/StopSec/StopSec-System}}.
\newblock
\newblock
\shownote{Accessed: 2025-07-01}.


\bibitem[Dogan-Tusha et~al\mbox{.}(2025)]%
        {dogantusha2025evaluation}
\bibfield{author}{\bibinfo{person}{Seda Dogan-Tusha}, \bibinfo{person}{Armed Tusha}, \bibinfo{person}{Muhammad~Iqbal Rochman}, \bibinfo{person}{Hossein Nasiri}, \bibinfo{person}{Joshua~Roy Palathinkal}, \bibinfo{person}{Mike Atkins}, {and} \bibinfo{person}{Monisha Ghosh}.} \bibinfo{year}{2025}\natexlab{}.
\newblock \showarticletitle{Evaluation of Indoor/Outdoor Sharing in the Unlicensed 6 GHz Band}.
\newblock \bibinfo{journal}{\emph{arXiv preprint arXiv:2505.18359}} (\bibinfo{year}{2025}).
\newblock


\bibitem[Foundation({[n.\,d.]})]%
        {apachebench}
\bibfield{author}{\bibinfo{person}{The Apache~Software Foundation}.} \bibinfo{year}{[n.\,d.]}\natexlab{}.
\newblock \bibinfo{title}{{ApacheBench}}.
\newblock
\newblock
\urldef\tempurl%
\url{https://httpd.apache.org/docs/current/programs/ab.html}
\showURL{%
\tempurl}
\newblock
\shownote{Accessed: 06.16.2025}.


\bibitem[Gaffney et~al\mbox{.}(2022)]%
        {gaffney2022sqlite}
\bibfield{author}{\bibinfo{person}{Kevin~P Gaffney}, \bibinfo{person}{Martin Prammer}, \bibinfo{person}{Larry Brasfield}, \bibinfo{person}{D~Richard Hipp}, \bibinfo{person}{Dan Kennedy}, {and} \bibinfo{person}{Jignesh~M Patel}.} \bibinfo{year}{2022}\natexlab{}.
\newblock \showarticletitle{SQLite: past, present, and future}.
\newblock \bibinfo{journal}{\emph{Proceedings of the VLDB Endowment}} \bibinfo{volume}{15}, \bibinfo{number}{12} (\bibinfo{year}{2022}).
\newblock


\bibitem[Golomb(1982)]%
        {GolombMLS}
\bibfield{author}{\bibinfo{person}{Solomon~W. Golomb}.} \bibinfo{year}{1982}\natexlab{}.
\newblock \bibinfo{booktitle}{\emph{Shift Register Sequences}}.
\newblock \bibinfo{publisher}{Aegean Park Press}.
\newblock


\bibitem[{Google Cloud}({[n.\,d.]})]%
        {Google2024SAS}
\bibfield{author}{\bibinfo{person}{{Google Cloud}}.} \bibinfo{year}{[n.\,d.]}\natexlab{}.
\newblock \bibinfo{booktitle}{\emph{Spectrum {A}ccess {S}ystem}}.
\newblock
\urldef\tempurl%
\url{https://cloud.google.com/spectrum-access-system/docs/}
\showURL{%
\tempurl}


\bibitem[Hassan et~al\mbox{.}({[n.\,d.]})]%
        {Intel2024AFC}
\bibfield{author}{\bibinfo{person}{Yaghoobi Hassan}, \bibinfo{person}{Cordeiro Carlos}, \bibinfo{person}{Arefi Reza}, {and} \bibinfo{person}{Horne David}.} \bibinfo{year}{[n.\,d.]}\natexlab{}.
\newblock \bibinfo{booktitle}{\emph{Spectrum Sharing Using Automated Frequency Coordination}}.
\newblock
\urldef\tempurl%
\url{https://www.intel.com/content/dam/www/central-libraries/us/en/documents/2022-12/spectrum-sharing-auto-frequency-coord-whitepaper.pdf}
\showURL{%
\tempurl}
\newblock
\shownote{White paper by Intel}.


\bibitem[Hipp(2020)]%
        {sqlite2020hipp}
\bibfield{author}{\bibinfo{person}{Richard~D Hipp}.} \bibinfo{year}{2020}\natexlab{}.
\newblock \bibinfo{title}{{SQLite}}.
\newblock
\newblock
\urldef\tempurl%
\url{https://www.sqlite.org/index.html}
\showURL{%
\tempurl}
\newblock
\shownote{Accessed: 06.16.2025}.


\bibitem[{IEEE}(2009)]%
        {ieee80211n}
\bibfield{author}{\bibinfo{person}{{IEEE}}.} \bibinfo{year}{2009}\natexlab{}.
\newblock \bibinfo{title}{IEEE Standard for Information Technology—Telecommunications and Information Exchange Between Systems—Local and Metropolitan Area Networks—Specific Requirements—Part 11: Wireless LAN MAC and PHY Specifications—Amendment 5: Enhancements for Higher Throughput}.
\newblock \bibinfo{howpublished}{IEEE Std 802.11n-2009}.
\newblock
\newblock
\shownote{Available at: \url{https://standards.ieee.org/standard/802_11n-2009.html}}.


\bibitem[{ISED Canada}(2022)]%
        {ISED2022Canada}
\bibfield{author}{\bibinfo{person}{{ISED Canada}}.} \bibinfo{year}{December 2022}\natexlab{}.
\newblock \bibinfo{title}{{Automated Frequency Coordination (AFC) System Specifications for the 6 GHz (5925-6875 MHz) Frequency Band}}.
\newblock
\newblock
\urldef\tempurl%
\url{https://ised-isde.canada.ca/site/spectrum-management-telecommunications/sites/default/files/attachments/2022/DBS-06-i1-2022-12EN.pdf}
\showURL{%
\tempurl}


\bibitem[{ITU Radiocommunication Sector}(1997)]%
        {itur1997guidelines}
\bibfield{author}{\bibinfo{person}{{ITU Radiocommunication Sector}}.} \bibinfo{year}{1997}\natexlab{}.
\newblock \bibinfo{booktitle}{\emph{Guidelines for Evaluation of Radio Transmission}}.
\newblock \bibinfo{type}{{T}echnical {R}eport}. \bibinfo{institution}{ITU-R}.
\newblock
\newblock
\shownote{M.1225}.


\bibitem[ITU-Report-2003(2003)]%
        {ITU-RA769-2}
\bibfield{author}{\bibinfo{person}{ITU-Report-2003}.} \bibinfo{year}{2003}\natexlab{}.
\newblock \showarticletitle{Protection criteria used for radio astronomical measurements}.
\newblock \bibinfo{journal}{\emph{Recommendation RA.769-2}} (\bibinfo{date}{March} \bibinfo{year}{2003}).
\newblock


\bibitem[ITU-Report-2012(2012)]%
        {ITU-RS2017-0}
\bibfield{author}{\bibinfo{person}{ITU-Report-2012}.} \bibinfo{year}{Aug. 2012}\natexlab{}.
\newblock \showarticletitle{Performance and interference criteria for satellite passive remote sensing}.
\newblock \bibinfo{journal}{\emph{Recommendation ITU-R RS.2017-0}} (\bibinfo{year}{Aug. 2012}).
\newblock


\bibitem[Leshem et~al\mbox{.}(1999)]%
        {Leshem1999detection}
\bibfield{author}{\bibinfo{person}{A. Leshem}, \bibinfo{person}{A.-J. van~der Veen}, {and} \bibinfo{person}{Ed. Deprettere}.} \bibinfo{year}{1999}\natexlab{}.
\newblock \showarticletitle{Detection and blanking of {GSM} interference in radio-astronomical observations}. In \bibinfo{booktitle}{\emph{1999 2nd IEEE Workshop on Signal Processing Advances in Wireless Communications (Cat. No.99EX304)}}. \bibinfo{pages}{374--377}.
\newblock
\urldef\tempurl%
\url{https://doi.org/10.1109/SPAWC.1999.783096}
\showDOI{\tempurl}


\bibitem[Lin and Costello(2004)]%
        {lin2004error}
\bibfield{author}{\bibinfo{person}{Shu Lin} {and} \bibinfo{person}{Daniel~J. Costello}.} \bibinfo{year}{2004}\natexlab{}.
\newblock \bibinfo{booktitle}{\emph{Error Control Coding: Fundamentals and Applications} (\bibinfo{edition}{2nd} ed.)}.
\newblock \bibinfo{publisher}{Prentice Hall}, \bibinfo{address}{Upper Saddle River, NJ, USA}.
\newblock


\bibitem[Marsh(2020)]%
        {marsh2020att}
\bibfield{author}{\bibinfo{person}{Joan Marsh}.} \bibinfo{year}{2020}\natexlab{}.
\newblock \bibinfo{title}{{AT\&T} Statement on {FCC} Order to Allow Unlicensed Devices in 6 {GHz} Band}.
\newblock
\newblock
\urldef\tempurl%
\url{https://www.attconnects.com/att-statement-on-fcc-order-to-allow-unlicensed-devices-in-6-ghz-band/}
\showURL{%
\tempurl}


\bibitem[Martikainen et~al\mbox{.}(2023)]%
        {Martikainen2023DSS}
\bibfield{author}{\bibinfo{person}{Henrik Martikainen}, \bibinfo{person}{Mikko Majamaa}, {and} \bibinfo{person}{Jani Puttonen}.} \bibinfo{year}{2023}\natexlab{}.
\newblock \showarticletitle{Coordinated Dynamic Spectrum Sharing Between Terrestrial and Non-Terrestrial Networks in 5G and Beyond}. In \bibinfo{booktitle}{\emph{2023 IEEE 24th International Symposium on a World of Wireless, Mobile and Multimedia Networks (WoWMoM)}}. \bibinfo{pages}{419--424}.
\newblock
\urldef\tempurl%
\url{https://doi.org/10.1109/WoWMoM57956.2023.00074}
\showDOI{\tempurl}


\bibitem[Mitić et~al\mbox{.}(2012)]%
        {Dragan2012calculating}
\bibfield{author}{\bibinfo{person}{Dragan Mitić}, \bibinfo{person}{Aleksandar Lebl}, {and} \bibinfo{person}{Zarko Markov}.} \bibinfo{year}{2012}\natexlab{}.
\newblock \showarticletitle{Calculating the required number of bits in the function of confidence level and error probability estimation}.
\newblock \bibinfo{journal}{\emph{Serbian Journal of Electrical Engineering}}  \bibinfo{volume}{9} (\bibinfo{year}{2012}), \bibinfo{pages}{361--375}.
\newblock


\bibitem[{National Research Council}(2010)]%
        {committee2010spectrum}
\bibfield{author}{\bibinfo{person}{{National Research Council}}.} \bibinfo{year}{2010}\natexlab{}.
\newblock \bibinfo{booktitle}{\emph{Spectrum Management for Science in the 21st Century}}.
\newblock \bibinfo{publisher}{National Academies Press}.
\newblock
\showISBNx{0-309-14686-0}


\bibitem[Palacios et~al\mbox{.}(2025)]%
        {palacios2025hidden}
\bibfield{author}{\bibinfo{person}{Ashton Palacios}, \bibinfo{person}{Daniel Harman}, \bibinfo{person}{Christopher Kitras}, \bibinfo{person}{Elle Kelsey}, \bibinfo{person}{Mitchell~C Burnett}, \bibinfo{person}{Willie~K Harrison}, {and} \bibinfo{person}{Philip Lundrigan}.} \bibinfo{year}{2025}\natexlab{}.
\newblock \showarticletitle{Hidden in Plain Sight: Communicating Using Interference}. In \bibinfo{booktitle}{\emph{2025 IEEE International Symposium on Dynamic Spectrum Access Networks (DySPAN)}}.
\newblock


\bibitem[Park et~al\mbox{.}(2024)]%
        {park2024AFC}
\bibfield{author}{\bibinfo{person}{Seungkeun Park}, \bibinfo{person}{Bongsu Kim}, \bibinfo{person}{Igor Kim}, {and} \bibinfo{person}{Jun-Bae Seo}.} \bibinfo{year}{2024}\natexlab{}.
\newblock \showarticletitle{{Deploying Automated Frequency Coordination System for WiFi 6E in South Korea: {C}hallenges and Opportunities}}.
\newblock \bibinfo{journal}{\emph{IEEE Communications Magazine}} \bibinfo{volume}{62}, \bibinfo{number}{1} (\bibinfo{year}{2024}), \bibinfo{pages}{112--118}.
\newblock
\urldef\tempurl%
\url{https://doi.org/10.1109/MCOM.001.2300446}
\showDOI{\tempurl}


\bibitem[Rice(2008)]%
        {rice2008digital}
\bibfield{author}{\bibinfo{person}{Michael~B. Rice}.} \bibinfo{year}{2008}\natexlab{}.
\newblock \bibinfo{booktitle}{\emph{Digital Communications: A Discrete-Time Approach}}.
\newblock \bibinfo{publisher}{Pearson Prentice Hall}, \bibinfo{address}{Upper Saddle River, NJ}.
\newblock
\showISBNx{9780130304971}


\bibitem[{RTCA Report}(2020)]%
        {RTCA2020FAA}
\bibfield{author}{\bibinfo{person}{{RTCA Report}}.} \bibinfo{year}{Oct. 07, 2020}\natexlab{}.
\newblock \bibinfo{title}{{Assessment of C-Band Mobile Telecommunications Interference Impact on Low Range Radar Altimeter Operations}}.
\newblock
\newblock
\urldef\tempurl%
\url{https://www.rtca.org/wp-content/uploads/2020/10/SC-239-5G-Interference-Assessment-Report_274-20-PMC-2073_accepted_changes.pdf}
\showURL{%
\tempurl}
\newblock
\shownote{RTCA Paper No. 274-20/PMC-2073}.


\bibitem[Salahdine and Kaabouch(2016)]%
        {salahdine2016matched}
\bibfield{author}{\bibinfo{person}{Fethi Salahdine} {and} \bibinfo{person}{Naima Kaabouch}.} \bibinfo{year}{2016}\natexlab{}.
\newblock \showarticletitle{Matched filter detection with dynamic threshold for cognitive radio networks}.
\newblock \bibinfo{journal}{\emph{arXiv preprint arXiv:1609.08398}} (\bibinfo{year}{2016}).
\newblock
\urldef\tempurl%
\url{https://arxiv.org/abs/1609.08398}
\showURL{%
\tempurl}


\bibitem[Sarbhai et~al\mbox{.}(2024a)]%
        {sarbhai2024reactive}
\bibfield{author}{\bibinfo{person}{Aarushi Sarbhai}, \bibinfo{person}{Frost Mitchell}, \bibinfo{person}{Sneha Kasera}, \bibinfo{person}{Aditya Bhaskara}, \bibinfo{person}{Jacobus Van~der Merwe}, {and} \bibinfo{person}{Neal Patwari}.} \bibinfo{year}{2024}\natexlab{a}.
\newblock \showarticletitle{Reactive Spectrum Sharing with Radio Dynamic Zones}. In \bibinfo{booktitle}{\emph{2024 IEEE International Symposium on Dynamic Spectrum Access Networks (DySPAN)}}. \bibinfo{pages}{429--438}.
\newblock


\bibitem[Sarbhai et~al\mbox{.}(2024b)]%
        {Aarushi2024RDZ}
\bibfield{author}{\bibinfo{person}{Aarushi Sarbhai}, \bibinfo{person}{Frost Mitchell}, \bibinfo{person}{Sneha Kasera}, \bibinfo{person}{Aditya Bhaskara}, \bibinfo{person}{Jacobus Van~der Merwe}, {and} \bibinfo{person}{Neal Patwari}.} \bibinfo{year}{2024}\natexlab{b}.
\newblock \showarticletitle{Reactive Spectrum Sharing with Radio Dynamic Zones}. In \bibinfo{booktitle}{\emph{2024 IEEE International Symposium on Dynamic Spectrum Access Networks (DySPAN)}}. \bibinfo{pages}{429--438}.
\newblock
\urldef\tempurl%
\url{https://doi.org/10.1109/DySPAN60163.2024.10632820}
\showDOI{\tempurl}


\bibitem[Sohul et~al\mbox{.}(2015)]%
        {Sohul2015SAS}
\bibfield{author}{\bibinfo{person}{Munawwar~M. Sohul}, \bibinfo{person}{Miao Yao}, \bibinfo{person}{Taeyoung Yang}, {and} \bibinfo{person}{Jeffrey~H. Reed}.} \bibinfo{year}{2015}\natexlab{}.
\newblock \showarticletitle{Spectrum access system for the citizen broadband radio service}.
\newblock \bibinfo{journal}{\emph{IEEE Communications Magazine}} \bibinfo{volume}{53}, \bibinfo{number}{7} (\bibinfo{year}{2015}), \bibinfo{pages}{18--25}.
\newblock
\urldef\tempurl%
\url{https://doi.org/10.1109/MCOM.2015.7158261}
\showDOI{\tempurl}


\bibitem[{University of Utah and the POWDER Project Team}(2023)]%
        {powder}
\bibfield{author}{\bibinfo{person}{{University of Utah and the POWDER Project Team}}.} \bibinfo{year}{2023}\natexlab{}.
\newblock \bibinfo{booktitle}{\emph{Platform for {O}pen {W}ireless {D}ata-driven {E}xperimental {R}esearch ({POWDER})}}.
\newblock
\urldef\tempurl%
\url{https://powderwireless.net/}
\showURL{%
\tempurl}


\bibitem[{US Federal Communications Commission}(2021a)]%
        {FCC2024AFC}
\bibfield{author}{\bibinfo{person}{{US Federal Communications Commission}}.} \bibinfo{year}{2021}\natexlab{a}.
\newblock \bibinfo{title}{{OET} Announces Approval of seven 6 {GH}z Band Automated Frequency Coordination Systems for Commercial Operation and Seeks Comment on {C3} Spectra’s Proposed {AFC} System}.
\newblock
\newblock
\urldef\tempurl%
\url{https://docs.fcc.gov/public/attachments/DA-24-166A1.pdf}
\showURL{%
\tempurl}
\newblock
\shownote{Public Notice ET Docket No. 21-352}.


\bibitem[{US Federal Communications Commission}(2021b)]%
        {FCC2022AFC}
\bibfield{author}{\bibinfo{person}{{US Federal Communications Commission}}.} \bibinfo{year}{2021}\natexlab{b}.
\newblock \bibinfo{title}{{OET announces conditional approval for 6 {GHz} band {AFC} systems}}.
\newblock
\newblock
\urldef\tempurl%
\url{https://docs.fcc.gov/public/attachments/DA-22-1146A1.pdf}
\showURL{%
\tempurl}
\newblock
\shownote{Public Notice ET Docket No. 21-352}.


\bibitem[{US Federal Communications Commission}(2023)]%
        {FCC2023TAC}
\bibfield{author}{\bibinfo{person}{{US Federal Communications Commission}}.} \bibinfo{year}{2023}\natexlab{}.
\newblock \bibinfo{title}{A Preliminary View of Spectrum Bands in the 7.125–24 {GHz} Range and a Summary of Spectrum Sharing Frameworks}.
\newblock \bibinfo{howpublished}{Technical Advisory Council White Paper}.
\newblock
\newblock
\shownote{US Federal Communications Commission}.


\bibitem[Vejlgaard et~al\mbox{.}(2017)]%
        {vejlgaard2017coverage}
\bibfield{author}{\bibinfo{person}{Benny Vejlgaard}, \bibinfo{person}{Mads Lauridsen}, \bibinfo{person}{Huan Nguyen}, \bibinfo{person}{Istv{\'a}n~Z Kov{\'a}cs}, \bibinfo{person}{Preben Mogensen}, {and} \bibinfo{person}{Mads Sorensen}.} \bibinfo{year}{2017}\natexlab{}.
\newblock \showarticletitle{Coverage and capacity analysis of {S}igfox, {LoRa}, {GPRS}, and {NB-IoT}}. In \bibinfo{booktitle}{\emph{2017 IEEE 85th Vehicular Technology Conference (VTC Spring)}}. \bibinfo{pages}{1--5}.
\newblock


\bibitem[Welch({[n.\,d.]})]%
        {welch2014florida}
\bibfield{author}{\bibinfo{person}{Chris Welch}.} \bibinfo{year}{[n.\,d.]}\natexlab{}.
\newblock \showarticletitle{Florida man drove around as a cellphone-jamming vigilante for two years}.
\newblock \bibinfo{journal}{\emph{The Verge}} (\bibinfo{year}{[n.\,d.]}).
\newblock
\urldef\tempurl%
\url{https://www.theverge.com/2014/5/1/5672762/man-faces-48000-fine-for-driving-with-cellphone-jammer}
\showURL{%
\tempurl}


\bibitem[Weldegebriel et~al\mbox{.}(2024)]%
        {Meles2024pseudonymetry}
\bibfield{author}{\bibinfo{person}{Meles~G. Weldegebriel}, \bibinfo{person}{Jie Wang}, \bibinfo{person}{Greg Hellbourg}, \bibinfo{person}{Ning Zhang}, {and} \bibinfo{person}{Neal Patwari}.} \bibinfo{year}{2024}\natexlab{}.
\newblock \showarticletitle{Watermarking of {OFDM} for {P}seudonymetry: {A}nalysis and {E}xperimental {R}esults}. In \bibinfo{booktitle}{\emph{2024 IEEE International Conference on Communications Workshops (ICC Workshops)}}. \bibinfo{pages}{317--322}.
\newblock
\urldef\tempurl%
\url{https://doi.org/10.1109/ICCWorkshops59551.2024.10615923}
\showDOI{\tempurl}


\bibitem[Weldegebriel et~al\mbox{.}(2022)]%
        {Meles2022pseudonymetry}
\bibfield{author}{\bibinfo{person}{Meles~G. Weldegebriel}, \bibinfo{person}{Jie Wang}, \bibinfo{person}{Ning Zhang}, {and} \bibinfo{person}{Neal Patwari}.} \bibinfo{year}{2022}\natexlab{}.
\newblock \showarticletitle{Pseudonymetry: Precise, Private Closed Loop Control for Spectrum Reuse with Passive Receivers}. In \bibinfo{booktitle}{\emph{2022 IEEE International Conference on RFID (RFID)}}. \bibinfo{pages}{91--96}.
\newblock
\urldef\tempurl%
\url{https://doi.org/10.1109/RFID54732.2022.9795976}
\showDOI{\tempurl}


\bibitem[Wetzel and Meyer(2004)]%
        {MIM-UMTS}
\bibfield{author}{\bibinfo{person}{Susanne Wetzel} {and} \bibinfo{person}{Ulrike Meyer}.} \bibinfo{year}{2004}\natexlab{}.
\newblock \showarticletitle{A Man-in-the-Middle Attack on {UMTS}}.
\newblock \bibinfo{journal}{\emph{WiSe '04: 3rd ACM workshop on Wireless security}} (\bibinfo{year}{2004}), \bibinfo{pages}{90--97}.
\newblock


\bibitem[{Wikipedia contributors}(2025)]%
        {TTLDef}
\bibfield{author}{\bibinfo{person}{{Wikipedia contributors}}.} \bibinfo{year}{2025}\natexlab{}.
\newblock \bibinfo{title}{Time to live}.
\newblock \bibinfo{howpublished}{\url{https://en.wikipedia.org/wiki/Time_to_live}}.
\newblock
\newblock
\shownote{Accessed: 2025-06-09}.


\bibitem[{Wireless Innovation Forum Spectrum Sharing Committee}(2018)]%
        {WInnForum_CBRSTechReq}
\bibfield{author}{\bibinfo{person}{{Wireless Innovation Forum Spectrum Sharing Committee}}.} \bibinfo{year}{2018}\natexlab{}.
\newblock \bibinfo{booktitle}{\emph{CBRS Technical Requirements (WInnForum Baseline Standards)}}.
\newblock \bibinfo{type}{{T}echnical {R}eport}. \bibinfo{institution}{Wireless Innovation Forum}.
\newblock
\urldef\tempurl%
\url{https://cbrs.wirelessinnovation.org/release-1-of-the-baseline-standard-specifications}
\showURL{%
\tempurl}
\newblock
\shownote{Baseline Standards Release 1 for CBRS}.


\bibitem[{Wireless Innovation Forum Spectrum Sharing Committee}(2022)]%
        {WInnForum_TS-0016}
\bibfield{author}{\bibinfo{person}{{Wireless Innovation Forum Spectrum Sharing Committee}}.} \bibinfo{year}{2022}\natexlab{}.
\newblock \bibinfo{booktitle}{\emph{Spectrum Access System (SAS) to CBSD Interface Technical Specification}}.
\newblock \bibinfo{type}{Technical Report WINNF-TS-0016v1.2.7}. \bibinfo{institution}{Wireless Innovation Forum}.
\newblock
\urldef\tempurl%
\url{https://winnf.memberclicks.net/assets/CBRS/WINNF-TS-0016.pdf}
\showURL{%
\tempurl}


\bibitem[{Wireless Spectrum R\&D Interagency WG, NITRD, NSTC}(2024)]%
        {NSF2024spectrum}
\bibfield{author}{\bibinfo{person}{{Wireless Spectrum R\&D Interagency WG, NITRD, NSTC}}.} \bibinfo{year}{2024}\natexlab{}.
\newblock \bibinfo{booktitle}{\emph{National Spectrum Research and Development Plan}}.
\newblock
\urldef\tempurl%
\url{https://www.whitehouse.gov/wp-content/uploads/2024/10/National-Spectrum-RD-Plan-2024.pdf}
\showURL{%
\tempurl}
\newblock
\shownote{National Science and Technology Council Working Group Report}.


\bibitem[Yang et~al\mbox{.}(2017)]%
        {yang2017narrowband}
\bibfield{author}{\bibinfo{person}{Wenjie Yang}, \bibinfo{person}{Mao Wang}, \bibinfo{person}{Jingjing Zhang}, \bibinfo{person}{Jun Zou}, \bibinfo{person}{Min Hua}, \bibinfo{person}{Tingting Xia}, {and} \bibinfo{person}{Xiaohu You}.} \bibinfo{year}{2017}\natexlab{}.
\newblock \showarticletitle{Narrowband wireless access for low-power massive internet of things: A bandwidth perspective}.
\newblock \bibinfo{journal}{\emph{IEEE Wireless Communications Magazine}} \bibinfo{volume}{24}, \bibinfo{number}{3} (\bibinfo{year}{2017}), \bibinfo{pages}{138--145}.
\newblock


\end{thebibliography}

\end{document}